\documentclass[12pt,amsmath,amssymb,longbibliography,nofootinbib,showkeys,eprint]{revtex4-1}
\usepackage{filecontents} 
\usepackage{epigraph}
\usepackage{graphicx}
\usepackage{epstopdf}
\usepackage{braket}
\usepackage{bm}
\usepackage{numprint}
\usepackage{float}
\usepackage[T1,T2A]{fontenc}
\usepackage[utf8]{inputenc}
\usepackage{xr}
\usepackage{amsthm}
\theoremstyle{definitions}

\newtheorem*{lem}{Lemma}
\usepackage{calc}
\usepackage{seqsplit}
\usepackage{hyperref}

\allowdisplaybreaks
\begin{document}
\preprint{V.M.}
\title{Market Dynamics:
  On Directional Information Derived From
  (Time, Execution Price, Shares Traded) Transaction Sequences.
}
\author{Vladislav Gennadievich \surname{Malyshkin}} 
\email{malyshki@ton.ioffe.ru}
\affiliation{Ioffe Institute, Politekhnicheskaya 26, St Petersburg, 194021, Russia}

\date{February, 15, 2018}

\begin{abstract}
\begin{verbatim}
$Id: DirectionalInformationFromTimePriceSharesTraded.tex,v 1.523 2019/05/01 20:55:53 mal Exp $
\end{verbatim}
A new approach to obtaining market--directional information, based on a non--stationary solution
to the dynamic equation
``future price tends to the value that maximizes the number of shares traded per unit time'' \cite{2015arXiv151005510G}
is presented. 
In our previous work\cite{2016arXiv160204423G}, we established
that it is the share execution flow ($I=dV/dt$) and not the share trading volume ($V$) that 
is the driving force of the market, and that asset prices are
much more sensitive to the execution flow $I$ (the dynamic impact) than to the traded volume $V$ (the regular impact).
In this paper, an important advancement is achieved: we define 
the ``scalp--price'' \hyperref[PPScalpedWH]{${\cal P}$} as the sum of only those price 
moves that are relevant to market dynamics;
\hyperref[WprojIH]{the criterion} of relevance is a high $I$.
Thus, only ``follow the market'' (and not ``little bounce'') events are included in ${\cal P}$.
Changes in the scalp--price defined this way indicate a market trend change --- not a bear market 
rally or a bull market sell--off;
the approach can be further extended to
\hyperref[FlPIHExampleFig]{non--local}
\hyperref[dptodFpIH]{price change}.
The software \hyperref[RunProgramUsage]{calculating}
the scalp--price given market observations triples
(time, execution price, shares traded) 
\href{http://www.ioffe.ru/LNEPS/malyshkin/AMuseOfCashFlowAndLiquidityDeficit.zip}{is available}
from the authors.
\end{abstract}

\maketitle

\section{\label{intro}Introduction}

Introduced in \cite{ArxivMalyshkinMuse},
the ultimate market dynamics problem --- finding evidence of existence (or 
proof of non--existence) of an automated trading machine
consistently making positive P\&L as a result of trading on a free market as an autonomous agent ---
can be formulated in its weak and strong forms:
\begin{itemize}
\item Weak form: Whether such an automated trading machine
  can exist at all using only legally available data. 
  (It can definitely
  exist in an illegal form -- e.g. when a brokerage uses client order flow information  to
  \href{https://en.wikipedia.org/wiki/Front_running}{frontrun}
  their own clients. This type of strategies typically rely on
  using proprietary information about clients' Supply--Demand future disbalance and on the subsequent monetization of 
  this information.)
\item Strong form: Whether such an automated trading machine
  can exist and be based solely on transaction sequences --
  say, the historical time series of (time, execution price, shares traded)
  market observations triples.  
  This information has supply and demand matched
  for every observation:
  at time $t$ trader A sold $v$ shares of some security at price $P$
   to trader B
  and received $v\cdot P$ dollars.
  Such a strategy can utilize only information
  about volume and execution flows.
\end{itemize}

We have shown  in \cite{2015arXiv151005510G,2016arXiv160204423G}
 that it is share execution flow $I=dV/dt$, not share trading volume $V$,
that is the driving force of the market
(see the Figs. 2 and 3 of Ref. \cite{2016arXiv160204423G}:
the asset price shows singularity at a high $I$,
but there is no price singularity at the maximal volume price,
the median of price--volume distribution).

In \cite{2015arXiv151005510G,2016arXiv160305313G},
the concept of liquidity deficit trading was introduced:
\textsl{open a position at low $I$, then
close already opened position at high $I$};
this is the only strategy that avoids catastrophic  P\&L losses.
This strategy is ideologically similar
to a classic volatility trading strategy: \textsl{buy a \href{https://en.wikipedia.org/wiki/Straddle}{straddle} at low volatility, sell it at high volatility, never go short volatility to avoid catastrophic  P\&L loss},
but is different from it by incorporating asset price directional contribution:
the decision is needed on whether
to open a long or a short position at low $I$.
In \cite{ArxivMalyshkinMuse}, the first attempt at finding
a non--stationary solution to the dynamic equation 
by linking asset price and liquidity deficit
via ``impact--from--the--future'' operator (adding
to execution flow a contribution
from not--yet--executed trades) was presented. In this paper, a different approach is developed.

Instead of adding not--yet--executed trades (impact--from--the--future),
we now consider removing from consideration
already executed trades (impact--from--the--past) corresponding to \textsl{high $I$ $\to$ low $I$} transitions.
A liquidity deficit trading strategy assumes that only \textsl{low $I$ $\to$ high $I$} transitions
will be captured by the trader.
The \textsl{high $I$ $\to$ low $I$} transitions are not to be used, as they are a major source of catastrophic risk.
A typical market behavior after a liquidity excess (high $I$)  event 
is to ``bounce a little,'' then go in the original direction of the market.
This creates an uncertainty of strategy. What does one bet on:  ``little bounce'' or ``follow the market''?
In contrast, after a liquidity deficit (low $I$)  event,
the market can \textsl{only} go in the direction of the market trend, eliminating this uncertainty.
This shows the importance of the assymetry of dynamic impact 
(price sensitivity to $I$ \cite{2016arXiv160204423G}): 
\textsl{low $I$ $\to$ high $I$} and \textsl{high $I$ $\to$ low $I$}
transitions are to be considered \textbf{separately}, as they lead to very different price behaviors.
This asymmetry is the topic of this study.
The scalp--function (\ref{WprojIH})
is introduced to comprise only those price moves relevant to market dynamics
(high $I$), which allows constructing scalp--price ${\cal P}$ (Fig. \ref{PPScalpedWH})
containing   only ``follow the market'' (and not ``little bounce'') events.
A change in the scalp--price indicates a market trend change,
not a bear market rally or a bull market sell--off.

\section{\label{mathapproach}Basic Mathematics}
The key  concept of
the
dynamic equation
``future price tends to the value that maximizes the number of shares traded per unit time''
\cite{2015arXiv151005510G,ArxivMalyshkinMuse}
is to find an averaging weight from the behavior of
a market dynamics operator $f$ (e.g. $dV/dt$, $V/t$, or $dI/dt$),
then to estimate some directional indicator (e.g. price change, signed volume, etc.) 
using the obtained weight.
Mathematically, the weight is considered in the form of an average depending on 
\href{https://en.wikipedia.org/wiki/Wave_function}{wavefunction} $\psi(x)=\sum_{k=0}^{n-1}\alpha_kQ_k(x)$:
$\psi^2(x(t))\omega(t)dt$,
an important generalization of
commonly--used parameter--independent fixed time scale averaging
such as the exponential
\href{https://en.wikipedia.org/wiki/Moving_average}{moving average}
: $\omega(t)dt$.
The bases $Q_m(x(t))\omega(t)dt$  we use in this paper are listed in Section \ref*{Mu-basis} of Ref. \cite{ArxivMalyshkinMuse}).
The problem is then reduced to a 
\href{http://www.netlib.org/lapack/lug/node54.html}{generalized eigenvalue problem}
of operator $\|f\|$:
\begin{eqnarray}
\Ket{f\Big|\psi_{f}^{[i]}}&=&\lambda_{f}^{[i]}\Ket{\psi_{f}^{[i]}} \label{GEVdef} \\
\sum\limits_{k=0}^{n-1} \Braket{Q_j|f|Q_k} \alpha^{[i]}_k &=&
  \lambda_f^{[i]} \sum\limits_{k=0}^{n-1} \Braket{ Q_j|Q_k} \alpha^{[i]}_k
  \label{GEV} \\ 
 \psi_{f}^{[i]}(x)&=&\sum\limits_{k=0}^{n-1} \alpha^{[i]}_k Q_k(x)
  \label{psiC}
\end{eqnarray}
The most general form of the averaging weight is a
\href{https://en.wikipedia.org/wiki/Density_matrix}{density matrix}:
\begin{eqnarray}
  \|\rho\|&=&\sum\limits_{i=0}^{n-1}\Ket{\psi_{\rho}^{[i]}}\lambda_{\rho}^{[i]}\Bra{\psi_{\rho}^{[i]}} \label{densmatrdef}\\
  f_{\rho}&=&\mathrm{Spur}\, \|f|\rho\|=\sum\limits_{i=0}^{n-1}\Braket{\psi_{\rho}^{[i]}|f|\psi_{\rho}^{[i]}}\lambda_{\rho}^{[i]}=
  \sum\limits_{i=0}^{n-1}\Braket{\psi_f^{[i]}|\rho|\psi_f^{[i]}}\lambda_f^{[i]}
  \label{densmatraver}
\end{eqnarray}
The most promising result of Refs. \cite{2015arXiv151005510G,ArxivMalyshkinMuse}
is averaging with the weight in the state $\Ket{\psi_I^{[IH]}}$ of the maximum
execution rate $I=dV/dt$ on the past sample.
This corresponds to the following density matrix and asset price:
\begin{eqnarray}
  \|\rho^{[IH]}\|&=&\Ket{\psi_I^{[IH]}}\Bra{\psi_I^{[IH]}}
  \label{rhoIH} \\
  p^{[IH]}&=&\Braket{\psi_I^{[IH]}|pI|\psi_I^{[IH]}}\Big/\lambda_I^{[IH]}
  \label{pIH}
\end{eqnarray}
Given a state $\Ket{\psi}$,
a number of values in this state can be calculated. Just a few examples.
Let's define
\begin{subequations}
  \label{inputmatricesAggregated}
\begin{align}
V_{s}(t)&=\int\limits_{t}^{t_{now}} p^s(t^{\prime}) dV^{\prime}
\label{cmdefV} \\
T_{s}(t)&=\int\limits_{t}^{t_{now}} p^s(t^{\prime}) dt^{\prime}
\label{cmdefT} 
  \end{align}
\end{subequations}
Here, $V_{0}(t)=V(t_{now})-V(t)$ is traded volume,
$V_{1}(t)$ is traded capital,
$V_{1}(t)/V_{0}(t)$ is volume--weighted average price,
$T_{0}(t)=t_{now}-t$,
and $T_{1}(t)/T_{0}(t)$ is time--weighted average price;
these are the values for the time interval:
between $t$ and $t_{now}$.
Then $p_{\{v,t\}}$ is \{volume,time\}--averaged price in the $\Ket{\psi}$ state,
$p_{\{V,T\}}$ is \{volume,time\} averaged aggregated price in the $\Ket{\psi}$ state,
calculated using the aggregated moments (\ref{inputmatricesAggregated}).
If $\Ket{\psi}$ is localized at some given $t$, then, approximately,
$p_{\{v,t\}}$ is the price at $t$ and $p_{\{V,T\}}$ are \{volume,time\}--weighted 
price moving average calculated for the time interval between $t$ and $t_{now}$:
\begin{subequations}
  \label{pvtdev:withoutSuccess}
\begin{align}
  p_v&=\frac{\Braket{\psi|pI|\psi}}{\Braket{\psi|I|\psi}} \label{pvDefine}\\
  p_t&=\frac{\Braket{\psi|p|\psi}}{\Braket{\psi|\psi}} \label{ptDefine} \\
  p_V&=\frac{\Braket{\psi|V_1|\psi}}{\Braket{\psi|V_0|\psi}} \label{pVDefine}\\
  p_T&=\frac{\Braket{\psi|T_1|\psi}}{\Braket{\psi|T_0\psi}} \label{pTDefine}
\end{align}
\end{subequations}
Moments $\Braket{Q_mV_s}$ and $\Braket{Q_mT_s}$
can be calculated from moments $\Braket{Q_m\,p^sI}$ and $\Braket{Q_m\,p^s}$
and, more generally, moments $\Braket{Q_m\frac{dF}{dt}}$ can be calculated from moments  $\Braket{Q_mF}$
 using integration by parts (see the Appendices
\ref{VMatrixElements} and \ref{dIdtMatrixElements} below).
In some cases, it is more convenient to directly integrate the wavefunction rather the individual basis functions as in (\ref{VbyParts}):
\begin{align}
  &w_{\psi}(t)=\int\limits_{-\infty}^{t}\psi^2(x(t^{\prime}))\omega(t^{\prime})dt^{\prime}
  =\omega(t)J(\psi^2(x(t))) \label{wfPartsdef} \\
  &\int\limits_{-\infty}^{t_{now}} F(t) \psi^2(x(t))\omega(t)dt
  = -\int\limits_{-\infty}^{t_{now}} \frac{dF}{dt} w_{\psi}(t)dt
  \label{wfPartsdefParts} \\
  &F(t_{now})=0: \quad \text{Boundary condition}
\end{align}
For the bases we use, $J(\cdot)$ in (\ref{wfPartsdef}) 
is analytically--known polynomial--to--polynomial mapping function\footnote{See
  the classes \texttt{\seqsplit{com/polytechnik/trading/\{WIntegratorLegendreShifted,WIntegratorLaguerre,WIntegratorMonomials\}.getPsi2WIntegratedDt()}}
  for numerical implementations)
}.
The Eq. (\ref{wfPartsdefParts}) allows
simultaneously 
to obtain the values of operator pairs: $(\|V_0\|,\|I\|)$, $(\|V_1\|, \|pI\|)$, etc.
in the state of a given $\psi(x)$.

What input data is required to obtain all the results
of this paper?
The $n\times n$ matrices $\Braket{Q_j|f|Q_k}$ ($j,k=[0 \dots n-1 ]$):
are calculated from generalized moments ($m=[0 \dots 2n-2 ]$):
\begin{subequations}
  \label{inputmoments}
  \begin{align}   
  &   \Braket{Q_m} \label{inputmoments:qm}\\
  &   \Braket{Q_m\,I} \label{inputmoments:Iqm} \\
    &   \Braket{Q_m\,pI} \\
    & \Braket{Q_m\frac{dp}{dt}}
  \end{align}
\end{subequations}
by applying basis functions multiplication operator (Eq. (\ref*{Mu-cmul}) of Ref. \cite{ArxivMalyshkinMuse}).
All the (\ref{inputmoments}) are calculated from
(Time, Price, Shares traded)
transaction sequence.

\section{\label{PnLoptimalPosition}P\&L and optimal position change}

Given a directional density matrix $\|\rho\|$,
how we do apply it?
A na\"{\i}ve answer is to 
average a directional attribute with it, for example:
\begin{itemize}
\item Use price change operator $\|f\|=\|\frac{dp}{dt}\|$
  (or $\|f\|=\|\frac{d^2p}{dt^2}\|$
  with some boundary condition from the Appendix \ref{dIdtMatrixElements}),
  calculate $\mathrm{Spur}\, \|f|\rho\|$; in a pure state $\|\rho\|=\Ket{\psi}\Bra{\psi}$,
 hereof $\mathrm{Spur}\, \|f|\rho\|=\Braket{\psi|f|\psi}$.
  Other directional
  attributes (signed volume, spread multiplied by signed volume,
  time difference spent in the order book, etc.) can be also considered\cite{2016arXiv160305313G}.
\item The state determining the dynamics
   often corresponds to a large $dI/dt$. Because $dI \approx I(t+dt)-I(t) >0$,
   $I=dV/dt$ is larger at the end of the interval.
  The asset price difference  $p_v-p_t$, with volume $dV$ and time $dt$ averaged in a state with such an asymmetry,
  is proportional to the directional component, where $p_v=\Braket{\psi|pI|\psi}/\Braket{\psi|I|\psi}$, and
  $p_t=\Braket{\psi|p|\psi}/\Braket{\psi|\psi}$.
  Note that such a difference between volume-- and time--averaged attribute
  $p_v-p_t$ carries directional information only
  in a state of large $dI/dt$, which makes an asymmetry of price averaging with 
  $dV$  and $dt$ correspond to $\delta p$. This is not the case
  in other states, e.g. trying to use the difference
  between volume-- and time--averaged price
  in the $\Ket{\psi_{I}^{[IH]}}$ state was fruitless in \cite{ArxivMalyshkinMuse}, see Appendix \ref{psiX} for a demonstration.
  It is now clear why: only the states with large $dI/dt$ provide
  weight asymmetry required to obtain directional information
  using $dV$ vs. $dt$ averaging.  
\end{itemize}

In \cite{2015arXiv151005510G} a P\&L operator
has been introduced in the Section II.E ``P\&L operator and trading strategy''.
Given a position change $dS$, the amount of shares
bought ($dS>0$) or sold ($dS<0$) during time interval $dt$,
the P\&L is\footnote{While the P\&L is $-\int pdS$, Eq. (\ref{PL}), the $\int IdS$, can be tried
as a directional indicator.}:
\begin{eqnarray}
  \mathrm{P\&L}&=-&\int p dS \label{PL} \\
  0&=&\int dS \label{PLconstraint}
\end{eqnarray}
The constraint (\ref{PLconstraint}) means:
total asset position should be zero in the
beginning and in the end of a trading period.
Formally, 
\begin{eqnarray}
  dS &=&\frac{d}{dt}\left(w(t)\frac{dp}{dt}\right) dt
  \label{formaldS}
\end{eqnarray}
where $w(t)$ is an arbitrary positive function,
provides
positive P\&L in (\ref{PL})
(integrate by parts and assume $\frac{dp}{dt}=0$ at the boundary
to satisfy (\ref{PLconstraint})).
Position increment $dS$ of optimal P\&L trading
has a symmetry of the second derivative
of price.
Note that in (\ref{formaldS})
other than $dp/dt$ attributes can be used,
designate it as ${\cal F}$,
for example: weighted price change ${\cal F}=\delta V\frac{dp}{dt}$ (price change multiplied by the volume traded
at this price), signed volume, signed volume multiplied by spread, etc.

There is a $dS$ answer of integral type:
  \begin{eqnarray}
    dS&=&\omega(t)\int\limits^{t}dt^{\prime}\int^{t^{\prime}}
    dt^{\prime\prime}\omega(t^{\prime\prime})p(t^{\prime\prime})
  \end{eqnarray}
  but it's non--local nature and the difficulty
  to choose integration limits to satisfy the 
  constraint (\ref{PLconstraint})
  make such an approach more difficult to implement.
  In the simplest form this approach
  is equivalent to
  buying below the median and selling above the median
  strategy considered in the Appendix \ref*{Mu-PnLTrading}
  of Ref. \cite{ArxivMalyshkinMuse}.

A very promising idea is a ``local trading strategy'' for $dS$ :
in $\Ket{\psi_I^{[IH]}}$ state  
buy at prices below the $p^{[IH]}$ from (\ref{pIH}), sell above the $p^{[IH]}$.
Corresponding $\|dS/dt\|$ operator is then:
\begin{eqnarray}
  dS&=& -\left(p-p^{[IH]}\right)dV  \label{dsEQppIH}\\
  \left\|\frac{dS}{dt}\right\|&=&
  -\left\|\left(p-p^{[IH]}\right)I\right\|
  \label{dSbuyBelowpIHsellAbovepIH} \\
  \mathrm{P\&L} &=& -\Braket{\psi_{I}^{[IH]}|p\frac{dS}{dt}|\psi_{I}^{[IH]}}
  =\Braket{\psi_I^{[IH]}|\left(p-p^{[IH]}\right)^2I|\psi_I^{[IH]}}
  \label{PdSbuyBelowpIHsellAbovepIH}
\end{eqnarray}
For this $dS$,
in the $\Ket{\psi_{I}^{[IH]}}$ state, the (\ref{PLconstraint})
condition is satisfied,
and the $\mathrm{P\&L}$ has a meaning of price standard deviation
(\ref{PdSbuyBelowpIHsellAbovepIH}).

\section{\label{BeyondTheWavefunction}Directional Information:
  Beyond The Wavefunction}
\begin{figure}[t]
  \includegraphics[width=16cm]{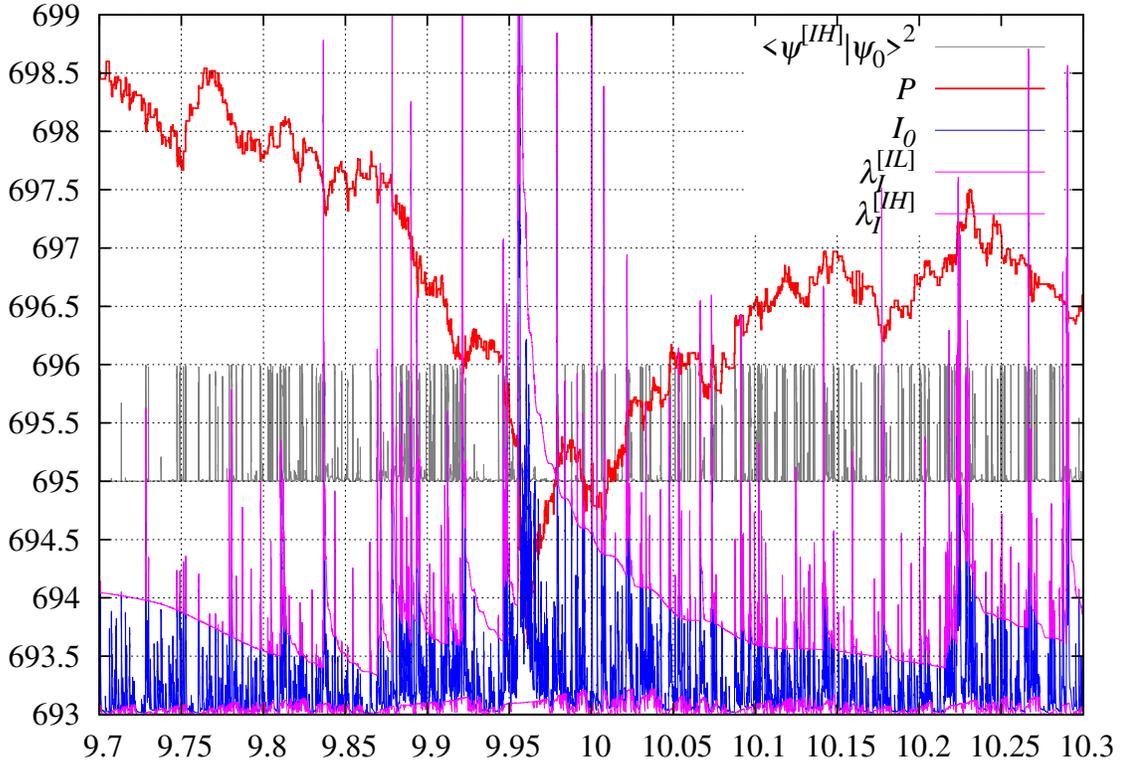}
  \caption{\label{ScalpedWH}
    The AAPL stock price on September, 20, 2012.
    The calculations in Shifted Legendre basis with $n=12$ and $\tau$=128sec. The $I_0$, $\lambda_I^{[IL]}$, $\lambda_I^{[IH]}$, and $\Braket{\psi_I^{[IH]}|\psi_0}^2$ projection (\ref{WprojIH}) are presented.   
  The execution flow $I$ is scaled and shifted to $693$, the projection
  is shifted to $695$ to fit the chart.
In between $[9.92\dots 9.94]$ the execution flow $I_0$ is small
and the $\Braket{\psi_I^{[IH]}|\psi_0}^2$ is close to zero,
thus make this interval non--contributing to scalp--moments.
What will happen to them,
when the $I_0=\Braket{\psi_0|I|\psi_0}$
is used as a scalp--function
instead of the $\Braket{\psi_I^{[IH]}|\psi_0}^2$?
In the $[9.92\dots 9.94]$ interval the $I_0$, while being small,
is not particularly zero and the contributions from this interval
will propagate to (\ref{dpdtAllW}); moreover the $I\to I+\mathrm{const}$
transform makes these contributions even larger. In the same
time the (\ref{WprojIH}) is almost zero in irrelevant to market dynamics
intervals and is invariant with respect to $I\to I+\mathrm{const}$
transform.
Effectively the $\Braket{\psi_I^{[IH]}|\psi_0}^2$
is the definition of scalp:
the condition of $I_0$ being high\cite{malyshkin2018spikes}.
  }
\end{figure}

As we have discussed in \cite{2015arXiv151005510G,2016arXiv160204423G}
the most interesting market behavior is observed at large $I$,
optimization problem $I=\frac{\Braket{\psi|I|\psi}}{\Braket{\psi|\psi}} \xrightarrow[{\psi}]{\quad }\max$
can be reduced to 
a generalized eigenvalue problem (\ref{GEV}) for $\|I\|$ operator:
\begin{eqnarray}
  \Ket{I\Big|\psi_{I}^{[i]}}&=&\lambda_{I}^{[i]}\Ket{\psi_{I}^{[i]}}
  \label{GEVI}
\end{eqnarray}
While the enter/exit conditions can be easily obtained from (\ref{GEVI}) as in (\ref{EVXData:I:fields}),
the directional information is a much trickier problem\cite{ArxivMalyshkinMuse}.
In \cite{2016arXiv160305313G}, the importance of P\&L dynamics was emphasized.
In Section \ref{PnLoptimalPosition} above, several trading strategies ($dS$),
retrospectively providing positive P\&L are presented. The goal, however,
is to build a strategy providing \textsl{future} positive P\&L.
Consider $p_t$ (\ref{ptDefine}) in the $\Ket{\psi_I^{[IH]}}$ state:
\begin{align}
  p_t^{[IH]}&=\frac{\Braket{\psi_I^{[IH]}|p|\psi_I^{[IH]}}}{\Braket{\psi_I^{[IH]}|\psi_I^{[IH]}}} \label{ptIHdef} \\
  P^{last}-p_t^{[IH]}&=\int dt\frac{dp}{dt}w_{\psi_I^{[IH]}}(t) \label{pLdiff}
\end{align}
The (\ref{pLdiff})
is just $dp/dt$ integration with the weigh (\ref{wfPartsdef})
for $\Ket{\psi_I^{[IH]}}$: the sum of the derivative values
with the proper weights give the last price minus the average.
The (\ref{pLdiff})
can be expressed via the $\Braket{Q_m \frac{dp}{dt}}$ moments
using an integration by parts of the Appendix \ref{VMatrixElements}.
The problem is reduced to calculation of the moments ($m=[0\dots 2n-1]$) from observations\footnote{
  Here the ``right'' sum is selected to simplify the recurrence
  by preserving the invariance of the time--grid.
  One can possibly use the ``middle'' sum with the $(t_l-t_{l-1})/2$ in the weight  $\omega(t)$
  and the basis $Q_m(x(t))$ functions argument.
  } sample $l=[1\dots M]$:
\begin{align}
  \Braket{Q_m \frac{dp}{dt}}&=\sum_{l=1}^{M} \left[p(t_l)-p(t_{l-1})\right]
   Q_m(x(t_l)) \omega(t_l)
\label{dpdtAll}
\end{align}
Then (\ref{dpdtAll})  can be substituted for
(\ref{pLdiff}) and the best directional answer of Ref. \cite{2015arXiv151005510G}: the last price minus the price in the $\Ket{\psi_I^{[IH]}}$ state
is obtained (the (\ref{pvDefine}) and (\ref{ptDefine}) are almost identical
in the $\Ket{\psi_I^{[IH]}}$ state).
These answers are the most general form that can be obtained using the ``pure wavefunction approach'':
all the answers are two quadratic forms ratio, possibly incoherently superposed
to a density matrix (\ref{densmatraver}).
However, as we have discussed above, ``not all observations are equal'':
only the events with a high $I$ are important for market dynamics.
Consider the  expression (\ref{pLdiff}) for a general attribute ${\cal F}$:
\begin{subequations}
  \label{dpW}
\begin{align}
  \mathrm{DIR\_scalped}&=\int dt {\cal F}(t)w_{\psi_I^{[IH]}}(t) \label{pLdiffW} \\
  \Braket{Q_m {\cal F}   }&=\sum_{l=1}^{M} (t_l-t_{l-1}){\cal F}_l  Q_m(x(t_l))\omega(t_l) 
  \label{dpdtAllW}
\end{align}
\end{subequations}
For
\begin{align}
  {\cal F}_l&=\frac{dp}{dt}=\frac{p(t_l)-p(t_{l-1})}{t_l-t_{l-1}}
  \label{Fleqdp}
\end{align}
the (\ref{dpW} is exactly the (\ref{pLdiff}) and (\ref{dpdtAll}).
Consider
\begin{align}
  {\cal F}(t)&=\frac{dp}{dt} {\cal S}(t) \label{FdP} \\
  (t_l-t_{l-1}){\cal F}_l&=\left(p(t_l)-p(t_{l-1})\right){\cal S}_l=\left(t_l-t_{l-1}\right)\frac{dp}{dt}{\cal S}_l \label{DirFact} \\
  {\cal S}(t)&:\text{$[0\dots 1]$ bounded function}\label{Wdef} \\
  {\cal S}_l&=\Braket{\psi_I^{[IH]}|\psi_0}^2 \text{\textbf{: For $t\in [-\infty \dots t_l]$ interval}} \label{WprojIH}
\end{align}
Now price change is multiplied by a $[0\dots 1]$
bounded scalp--function ${\cal S}(t)$ to select
``the relevance to market dynamics'' of any single observation moment $t_l$.
This way, we can remove from consideration all ``irrelevant''
observations, as discussed in the introduction;
the relevance is determined by estimating whether the current
execution flow $I_0$ is extremely large.
The answer obtained in  \cite{2015arXiv151005510G,malyshkin2018spikes} is:
for every $t_l$ observation solve the (\ref{GEVI})
problem for the interval $[-\infty \dots t_l]$
and consider the projection (\ref{WprojIH})
for time--shifted ($t_{now}=t_l$) problem (\ref{GEVI}).
The calculations are straightforward.
At time ``now,'' look back at all $[1\dots M]$ market observations,
calculate the sum (\ref{dpdtAllW}); for every term at $t_l$
also ``look back'' to construct a separate set of matrices $\Braket{Q_j|Q_k}$
and $\Braket{Q_j|I|Q_k}$ for the interval $l^{\prime}=[1\dots l]$
and calculate the scalp--function ${\cal S}$ from (\ref{WprojIH}).
This is a problem of $O(M^2)$ complexity
when approached directly, but it can be optimized
using recurrence relation for the moments calculated for different
observation intervals $l^{\prime}=[0\dots l]$, $l=[0\dots M]$.
The major difference with the (\ref{pLdiff})
is that the averaging (\ref{pLdiffW})
can no longer be written in the
density--matrix form (\ref{densmatraver})
with the original $\Braket{Q_m\frac{dp}{dt}}$ moments. The integration
weight in (\ref{pLdiff})
is obtained from the integration of (\ref{wfPartsdef}).
Using Theorem 3 from the Appendix A
of Ref. \cite{ArxivMalyshkinLebesgue},
any polynomial $P(x)$ of $2n-2$ degree can be
isomorphly mapped to a linear operator of the dimension $n$,
thus the density matrix, corresponding to the $w_{\psi_I^{[IH]}}$ averaging (\ref{pLdiff}),
can be readily obtained.
This is no longer the case for (\ref{pLdiffW}) averaging.
The scalp--function ${\cal S}$, while is easy to calculate
numerically, does not allow to reduce (\ref{pLdiffW}) averaging
to a density matrix  averaging (\ref{densmatraver})
of the original moments (\ref{dpdtAll});
we now need the scalp--moments (\ref{DirFact})
to average them with the $w_{\psi_I^{[IH]}}$.
This is similar to \href{https://en.wikipedia.org/wiki/Bloch_wave}{Bloch wavefunction} in quantum mechanics,
where the ``true'' wavefunction is considered as a product
of slow  and fast 
oscillating terms.
Now we have a product of slow $w_{\psi}(t)$
and fast ${\cal S}(t)$
changing weights in (\ref{pLdiffW}).
The greatest advantage of such a transition from
regular to scalp--moments,
is that the averaging weight can be very sharp.
Compare the $I_0$ in Fig. \ref{ScalpedWH}
with, calculated from the (\ref{inputmoments}) input
at fixed $t_{now}$,
the ``interpolated'' $I(y(-\infty\le t\le t_{now}))$  
in Fig. \ref{psiback}
of the Appendix \ref{psiX}: even for the dimension $n=12$ obtained
wavefunction states are not sufficiently localized
to select the sharp spikes in price changes
at high $I$. In the same time the dimension $n=12$
is perfectly OK for the execution flow $I$.
The scalp--function (\ref{Wdef}) is a practical way to unify
price and execution--flow dynamics within a single framework.

In the Fig. \ref{ScalpedWH}  the $\Braket{\psi_I^{[IH]}|\psi_0}^2$
projection (\ref{WprojIH})
along with $I_0$ and $\lambda_I^{[IH]}$ are presented.
One can clearly see that the (\ref{WprojIH}) is a very
good indicator of market activity,
the effect we have
noticed back in \cite{2015arXiv151005510G,malyshkin2018spikes}.
Now, however, we know how to apply this knowledge:
the criterion 
of current execution flow
being extremely high (such as $\Braket{\psi_I^{[IH]}|\psi_0}^2$)
can be  used as a scalp--function ${\cal S}$
when calculating the $dp/dt$ moments in (\ref{dpdtAllW}):
 multiply each $p(t_l)-p(t_{l-1})$  by the scalp--function.
This way only the relevant (high $I$) market moves will be accounted
in the scalp--moments $\Braket{ Q_m{\cal F}}=\Braket{ Q_m {\cal S}\frac{dp}{dt}}$.
Typical
\href{https://en.wikipedia.org/wiki/Scalping_(trading)}{scalping}
is price spikes (relatively some ``average''--like level)
identification technique
along with a set of rules to enter a trade and to take a profit/stoploss.
As we have shown\cite{2016arXiv160204423G}
the spikes in the execution flow, not in the price,
are responsible for market dynamics.

The (\ref{dpdtAllW}) main idea is to accumulate,
with the $Q_m(x(t))\omega(t)dt$ weight, a directional attribute,
such as $p(t_l)-p(t_{l-1})$,
 (Ref. \cite{2015arXiv151005510G} result)
\textsl{multiplied} by a scalp--function,
such as (\ref{WprojIH}) (this paper result);
in practice this is just a directional attribute transform (\ref{dptodF}).
Algorithmically, we need to listen for all trading events,
and, for each coming event in sequence,
obtain a directional attribute ${\cal F}_l$ from
the regular moments,
then calculate (\ref{dpdtAllW}) scalp--moments (recurrent optimization
make it very efficient computationally)
to obtain the directional information (\ref{pLdiffW}).
This can be implemented in several ways:
\begin{itemize}
\item Tick trading.
  As a transaction sequence consider every tick (execution or even limit order book event).
  For every tick $l$ calculate\footnote{
    Most of ${\cal D}_l=p(t_l)-p(t_{l-1})=0$ as most trading occur at the same price.
    Also note that $p_M-p_{m-1}=\sum_{l=m}^M{\cal D}_l$.
    For a different weight in the sum obtain (\ref{pLdiff}).
  }
\begin{align}  
  {\cal F}_l&=\frac{p(t_l)-p(t_{l-1})}{t_l-t_{l-1}}\Braket{\psi_I^{[IH]}|\psi_0}^2
  \label{DirFactSimple}
\end{align}
to obtain the ``filtered by relevance'' moments in (\ref{dpdtAllW}).
\item Assuming we have all the ticks data\footnote{
In practice, for US equity market, a sub--millisecond data can be obtained
at reasonable cost.
 For other markets, such as fixed income,
every tick data cannot be practically obtained.
Even for currency trading the fragmentation of the markets
along with prohibitively high prices on sub--millisecond data,
make any tick--trading approach practically unfeasible.
However, as we have discussed in Ref. \cite{ArxivMalyshkinMuse},
the time scale of market opportunities (along with liquidity available!)
expand well beyond sub--millisecond time scale,
maximal scale is determined by the availability of
high enough  fluctuations in the execution flow $I$,
at least an order of magnitude in $\lambda_I^{[IH]}/\lambda_I^{[IL]}$.
} ,
  instead of the price difference  some average
  multiplied by the scalp--function can be used:
\begin{align}
  {\cal F}_l&=\Braket{\psi_0|\frac{dp}{dt}|\psi_0}
  \Braket{\psi_I^{[IH]}|\psi_0}^2
  \label{DirFactProxy}
\end{align}
The (\ref{DirFactProxy}) uses the $\Ket{\psi_0}$ for the
interval $t\in[-\infty\dots t_l]$ with $t_{now}=t_l$.
The $\psi_0$ from (\ref{psiRN}) has an
internal time scale $1/\psi^2_0(x_0)$ (which is determined by the basis dimension $n$ and scale $\tau$), thus in (\ref{DirFactProxy}) the $dp/dt$ 
is averaged over the time $1/\psi^2_0(x_0)$.
The result is very similar to price tick (\ref{DirFactSimple}) approach,
see the Fig. \ref{PPCompare} below.
A quite similar result can also be obtained with
\begin{align}
  {\cal F}_l&=\psi_0^2(x_0)\left[\frac{\Braket{\psi_0|pI|\psi_0}}{\Braket{\psi_0|I|\psi_0}}
  -\frac{\Braket{\psi_0|p|\psi_0}}{\Braket{\psi_0|\psi_0}}\right]
  \Braket{\psi_I^{[IH]}|\psi_0}^2
  \label{DirPiPtroxy}
\end{align}
this corresponds to described above approach 
of the difference between
volume and time averaged price.
\item The (\ref{DirFactSimple}) and (\ref{DirFactProxy})
are  calculated in the $\Ket{\psi_0}$ state.
One can consider other states, the $\Ket{\psi_I^{[IH]}}$ is of special
interest
\begin{subequations}
  \label{momaltern}
\begin{align}
  {\cal F}_l&=\Braket{\psi_I^{[IH]}|\frac{dp}{dt}|\psi_I^{[IH]}}\Braket{\psi_I^{[IH]}|\psi_0}^2
  \label{dpdtnonlocal}\\
   {\cal F}_l&=2\left[\Braket{\psi_I^{[IH]}|pI|D(\psi_I^{[IH]})}-\Braket{\psi_I^{[IH]}|pI|\psi_I^{[IH]}}\Braket{\psi_I^{[IH]}|D(\psi_I^{[IH]})}
   \right]
   \label{DirpIProxy} \\
         {\cal F}_l&=\Braket{\psi_I^{[IH]}|\frac{d}{dt}pI|\Braket{\psi_I^{[IH]}}} \nonumber \\
         &=2\left[P^{last}\lambda_I^{[IH]}\Braket{\psi_I^{[IH]}|D(\psi_I^{[IH]})}
    -\Braket{\psi_I^{[IH]}|pI|D(\psi_I^{[IH]})}
    \right]\label{DirpIProxydPi}
\end{align}
\end{subequations}
An important feature of (\ref{momaltern}) is that
some of these ${\cal F}_l$ expressions (\ref{DirpIProxy}) and (\ref{DirpIProxydPi}) 
are calculated from $\|pI\|$ operator variation and have: 1. the dimension of capital
 2. the $\Braket{\psi_I^{[IH]}|\psi_0}^2$ factor
entering
due to the identity
\begin{align}
         \left(\psi_I^{[IH]}(x_0)\right)^2&=2\Braket{\psi_I^{[IH]}|D(\psi_I^{[IH]})}=
         \Braket{\psi_I^{[IH]}|\psi_0}^2\psi^2_0(x_0)
         \label{psix02}
\end{align}

\item An ability to use an expression,
  calculated from the regular moments $\Braket{\cdot}$, such as (\ref{momaltern})
  is a very important generalization of price change directional attribute (\ref{Fleqdp}).
  The minimal time--scale of such an attribute is $1/\psi_0^2(x_0)$,
  and the experiment shows that (\ref{DirFactSimple}) and (\ref{DirFactProxy})
  produce very similar results. This makes promising
  to consider a directional attributes of more general form in the $\Ket{\psi_0}$  state:
  calculate the ${\cal F}_l$, and use it as it were
  a regular price change. All the previously considered ${\cal F}_l$ were
  some kind a price change analogue. In Ref. \cite{ArxivMalyshkinMuse}
  two new directional attributes have been introduced:
  skewness and probability correlation.
  Consider the skewness (Eq. (\ref*{Mu-skewness}) of Ref. \cite{ArxivMalyshkinMuse})
  calculated out of four input moments:
\begin{align}
  \pi_s&=\Braket{\psi_0|p^sI|\psi_0} \label{piMoments0} \\
  s&=0,1,2,3 \nonumber \\
  \widetilde{\Gamma}&=\frac{2\overline{p}-p_{\min}-p_{\max}}
            {p_{\min}-p_{\max}} \label{skewnessG}
\end{align}
The idea is to build two--point Gauss quadrature
(the $p_{\min}$, $p_{\max}$ are min/max nodes of the quadrature,
Eq. (\ref*{Mu-evp3nodes}) of Ref. \cite{ArxivMalyshkinMuse},
and $\overline{p}=\pi_1/\pi_0$)
then to consider it's weight asymmetry as the asymmetry of the distribution.
The weigh asymmetry (\ref{skewnessG}) is actually
proportional to the difference between the median estimator $(p_{\min}+p_{\max})/2$
and the average $\overline{p}$. One can use the skewness
\begin{align}
  {\cal F}_l&=
  \left[p_{\max}-p_{\min}\right]\widetilde{\Gamma}\Braket{\psi_I^{[IH]}|\psi_0}^2
  \label{Flskewness}
\end{align}
as a directional attribute instead of price change.
The (\ref{Flskewness}) is calculated from the regular moments
$\Braket{Q_k}$, $\Braket{IQ_k}$, $\Braket{pIQ_k}$, $\Braket{p^2IQ_k}$, and $\Braket{p^3IQ_k}$,
then the ${\cal F}_l$ is used \textsl{as it were observed} at $t=t_l$.
This way we substitute price change by the skewness
calculated at $1/\psi_0^2(x_0)$ scale. The scalp--function
$\Braket{\psi_I^{[IH]}|\psi_0}^2$ makes only relevant to market dynamics
observations to contribute.

\item Variate the $\|pI\|$ in the $\Ket{\psi_I^{[IH]}}$
  state\footnote{The expression has the meaning of capital change
    due to   (\ref{psix02}) identity. For single asset consideration
    it is convenient to divide (\ref{DirVarPsi0AZavor1e6})
    by $\lambda_I^{[IH]}$.
  }
  with $\Ket{\psi_0}$:
\begin{align}  
    {\cal F}_l&=2\left[
      \Braket{\psi_I^{[IH]}|pI|\psi_0}-\Braket{\psi_I^{[IH]}|pI|\psi_I^{[IH]}}
      \Braket{\psi_I^{[IH]}|\psi_0}\right]\Braket{\psi_I^{[IH]}|\psi_0}
    \label{DirVarPsi0AZavor1e6}
\end{align}
If $\Ket{\psi_0}$ is the $\Ket{\psi_I^{[IH]}}$ then
(\ref{DirVarPsi0AZavor1e6}) is zero and no directional information is available.
The factor $\Braket{\psi_I^{[IH]}|\psi_0}$, which does not enter the (\ref{rqD1}),
is included in (\ref{DirVarPsi0AZavor1e6}) as a scalp--function; this factor also provides
proper sign invariance for $\psi\to-\psi$ transform: $\Braket{\psi_I^{[IH]}|\psi_0}$ is not squared
as it is in (\ref{WprojIH}).
One can also use a higher degree of $\Braket{\psi_I^{[IH]}|\psi_0}$
factor in  (\ref{DirVarPsi0AZavor1e6})
to make the peaks sharper.
\item Similar to (\ref{DirPiPtroxy}), but with $\Ket{\psi_I^{[IH]}}$ and $\Ket{\psi_0}$
\begin{align}
    {\cal F}_l&=\psi_0^2(x_0)\left[\frac{\Braket{\psi_0|pI|\psi_0}}{\Braket{\psi_0|I|\psi_0}}
    -\frac{\Braket{\psi_I^{[IH]}|pI|\psi_I^{[IH]}}}{\Braket{\psi_I^{[IH]}|I|\psi_I^{[IH]}}}
  \right]
  \Braket{\psi_I^{[IH]}|\psi_0}^{2}
  \label{DirP0PiH}
\end{align}
can be considered.
This is price difference in $\Ket{\psi_0}$ and $\Ket{\psi_I^{[IH]}}$ states.
Were it not for the scalp--function  $\Braket{\psi_I^{[IH]}|\psi_0}^2$
this would be almost Ref. \cite{2015arXiv151005510G} answer:
the difference between the last price and $p^{[IH]}$ (\ref{pIH}).
The scalp--function  makes this difference to be accumulated
only for the events of extremeny high $I_0$.
The (\ref{DirVarPsi0AZavor1e6}) and (\ref{DirP0PiH})
are zero
for $\Ket{\psi_I^{[IH]}}$  being equal to $\Ket{\psi_0}$,
thus satisfy  Ref. \cite{ArxivMalyshkinMuse}
Eq. (\ref*{Mu-dIfuturedir})
condition of ``no directional information
about the future available''.

\item
  All the ${\cal F}_l$ considered above are some kind of price change.
  Tick trading (\ref{DirFactSimple}) is last price minus previous price,
  the other (e.g. (\ref{DirFactProxy}), (\ref{DirPiPtroxy}), etc.)
  are calculated from the regular moments.
  It is a promising path  to combine tick and moments approaches.
\begin{align}
  {\cal F}_l&=\left[p_l-p^*_l\right]\Braket{\psi_I^{[IH]}|\psi_0}^2
  \label{FlFrpmPlPs} \\
  p^*_l&: \text{calculated from $\Braket{IQ_k}$ and $\Braket{pIQ_k}$
    moments}
\end{align}
Estimating the $p^*_l$
as (\ref{pIH}), or (\ref{pvtdev:withoutSuccess}) in the state
$\Ket{\psi_0}$ or $\Ket{\psi_I^{[IH]}}$
will not provide a good answer, the (\ref{momaltern}) is a demonstration.
A promising approach is to consider skweness like answer (\ref{skewnessG}).
Take (\ref{Flskewness}) but consider it in a different basis of dimension two,
replace  the basis $1$, $p(t)$ by $\Ket{\psi_0}$, $\Ket{\psi_I^{[IH]}}$,
as these are the states that are localized and relevant to market dynamics:
\begin{align}
&\left(
  \begin{array}{rr}
    \Braket{\psi_0|pI|\psi_0} & \Braket{\psi_0|pI|\psi_I^{[IH]}} \\
    \Braket{\psi_I^{[IH]}|pI|\psi_0} & \Braket{\psi_I^{[IH]}|pI|\psi_I^{[IH]}}
  \end{array}
  \right)
  \left(\begin{array}{l}
    \alpha_{0}^{[0,1]}\\
    \alpha_{1}^{[0,1]}
  \end{array}
  \right)
   = \nonumber \\
  &\qquad\qquad=\lambda_{p^*}^{[0,1]}
   \left(\begin{array}{rr}
    \Braket{\psi_0|I|\psi_0} & \Braket{\psi_0|I|\psi_I^{[IH]}} \\
    \Braket{\psi_I^{[IH]}|I|\psi_0} & \Braket{\psi_I^{[IH]}|I|\psi_I^{[IH]}}
  \end{array}
  \right)
  \left(\begin{array}{l}
    \alpha_{0}^{[0,1]}\\
    \alpha_{1}^{[0,1]}
  \end{array}
  \right)
  \label{lamevP}
\end{align}
The $\lambda_{p^*}^{[0]}$ and $\lambda_{p^*}^{[1]}$
eigenvalues give the  min/max price estimates,
that can be obtained in a state of $\Ket{\psi_0}$ and $\Ket{\psi_I^{[IH]}}$ superposition.
An answer similar to the skewness (\ref{Flskewness}) can be used as an estimator
of $p_l$ being low/high:
\begin{align}
  {\cal F}_l&=-z
  D_p
  \Braket{\psi_I^{[IH]}|\psi_0}^2
  \label{FlSkewnessPld2} \\
  D_p&=\frac{2p_l-\lambda_{p^*}^{[0]}+ \lambda_{p^*}^{[1]}}{\lambda_{p^*}^{[0]}- \lambda_{p^*}^{[1]}}
  \label{SkewnessPld2} \\
  z&=\begin{cases}
  \frac{|p(t_l)-p(t_{l-1})|}{t_l-t_{l-1}}\\
  \frac{V(t_l)-V(t_{l-1})}{t_l-t_{l-1}}\\
  \lambda_{p^*}^{[1]}- \lambda_{p^*}^{[0]}\\
  \dots
  \end{cases} \nonumber
\end{align}
The ${\cal F}_l$ is proportional to the difference between $p_l$
and $\frac{1}{2}\left[\lambda_{p^*}^{[0]}+ \lambda_{p^*}^{[1]}\right]$
is similar to Eq. (\ref*{Mu-skewnesslikeS0}) of Ref. \cite{ArxivMalyshkinMuse}.
This is an  approach generalizing  tick and moments approaches.
However, now the (\ref{SkewnessPld2}) is no longer $[-1:1]$ bounded
(it would be if one replaces $p_l$ by $\Braket{\psi_0|pI|\psi_0}\Big/\Braket{\psi_0|I|\psi_0}$).
A moments--only answer (without the last price used explicitly) can be also obtained:
\begin{align}
  {\cal F}_l&=z
  D_p
       \Braket{\psi_I^{[IH]}|\psi_0}^2
       \label{FlProbabilityCorrelation2DirSignedScalp} \\
       D_p&=  \frac{\left[\phi^{[1]}(x_0)\right]^2-\left[\phi^{[0]}(x_0)\right]^2}
       {\left[\phi^{[1]}(x_0)\right]^2+\left[\phi^{[0]}(x_0)\right]^2}
       \label{ProbabilityCorrelation2Dp}\\
       \phi^{[0,1]}(x)&=\alpha_0^{[0,1]}\psi_0(x)+\alpha_1^{[0,1]}\psi_I^{[IH]}(x)
       \label{phifromlam2} \\
z&=\begin{cases}
  \frac{|p(t_l)-p(t_{l-1})|}{t_l-t_{l-1}}\\
  \frac{V(t_l)-V(t_{l-1})}{t_l-t_{l-1}}\\
  \lambda_{p^*}^{[1]}- \lambda_{p^*}^{[0]}\\
  \dots
  \end{cases} \nonumber
\end{align}
 The sign of (\ref{FlProbabilityCorrelation2DirSignedScalp}) is determined by which one
of (\ref{lamevP}) eigenfunctions $\phi^{[0,1]}(x)$ is greater at $x_0$,
(\ref*{Mu-projdiff}) distance from Ref. \cite{ArxivMalyshkinMuse}.
The  directional factor $\frac{\left[\phi^{[1]}(x_0)\right]^2-\left[\phi^{[0]}(x_0)\right]^2}
{\left[\phi^{[1]}(x_0)\right]^2+\left[\phi^{[0]}(x_0)\right]^2}$
can be considered as probability correlation (Appendix \ref*{Mu-twoPvarcorrela}
of Ref. \cite{ArxivMalyshkinMuse})
between price and ``distance to now''.
In (\ref{FlSkewnessPld2}) and (\ref{FlProbabilityCorrelation2DirSignedScalp})
the scale factor 
$\left[\lambda_{p^*}^{[1]}- \lambda_{p^*}^{[0]}\right]$
by $D_p$ is replaced by more general form $z$,
what makes the scalp--price to preserve
the singularities for a variety of $D_p$ used.

\end{itemize}
As we have discussed in \cite{2015arXiv151005510G,2016arXiv160305313G},
price and price changes are secondary to execution flow and cannot
be used to determine market direction for the reason of insufficient
information. The main idea behind the scalp--moments is to replace in the sum (\ref{dpdtAll})
\begin{align}
  p(t_l)-p(t_{l-1})&\to (t_l-t_{l-1}){\cal F}_l  \label{dptodF} \\
  {\cal P}(t_{M})&=\sum\limits_{l=1}^{M}(t_l-t_{l-1}){\cal F}_l \label{PriceFromdF} \\
  \mathrm{DIR\_scalped}&={\cal P}^{last}-\Braket{\psi_I^{[IH]}|{\cal P}|\psi_I^{[IH]}}=
  \int dt {\cal F}(t)w_{\psi_I^{[IH]}}(t) \label{pLdiffWSum}
\end{align}
where ${\cal F}_l$ contains not only price changes, but also execution flow information.
A good ${\cal F}_l$ selection allows us to accumulate
much more directional information in the scalp--moments $\Braket{Q_m {\cal F} }$
compared to the information in the regular moments $\Braket{Q_m \frac{dp}{dt}}$.
If one sum all the ${\cal F}_l$ terms,
the ${\cal P}$, a generalized price can be obtained (\ref{PriceFromdF}).
The ${\cal P}$ is defined within a constant (it is convenient to take the last ``price''
${\cal P}^{last}$ equals to zero).
The transition from price $p$ to the scalp--price ${\cal P}$
makes all directional singularities  expressed much more clearly.
The directional information (\ref{pLdiffW}) now take the (\ref{pLdiffWSum}) form,
that is identical to (\ref{pLdiff}), but instead of price $p$ the scalp--price
${\cal P}$ is used.
If a trader wants to watch the prices --- he should be watching the scalp--price ${\cal P}$,
a much more informative characteristic in terms or market trend, than the regular price $p$.

\subsection{A Demonstration of Scalp--Price ${\cal P}$ Behavior}
\begin{figure}[t]
  \includegraphics[width=16cm]{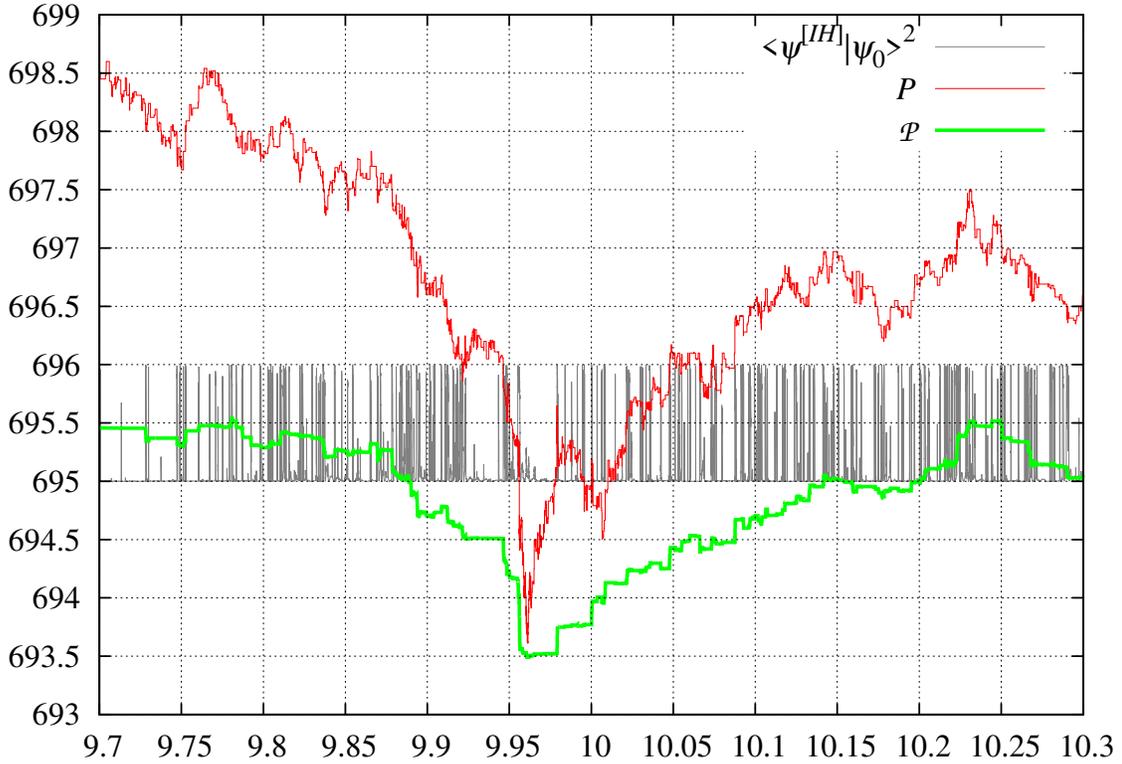}
  \caption{\label{PPScalpedWH}
    Price and scalp--price ${\cal P}$ for ${\cal F}_l$ from (\ref{DirFactProxy})
    are presented. The $\Braket{\psi_I^{[IH]}|\psi_0}^2$ is used as
    a scalp--function ${\cal S}(t)$ (\ref{Wdef}). Scalp--price is shifted
    to fit the chart. See Appendix \ref{CallAMusestucture} for data
    fields \texttt{T}, \texttt{p\_last}, \texttt{shares}, \texttt{p\_IH},
    \texttt{I.wH\_squared}, and 
    \texttt{getSumFdt()}
    corresponding to: time, price, shares traded, $p^{[IH]}$ (\ref{pIH}),
    scalp--function (\ref{WprojIH}),
    and scalp--price ${\cal P}$ (\ref{PriceFromdF}).
  }
\end{figure}

Before we go any further,
let us demonstrate scalp--price (\ref{PriceFromdF}) ${\cal P}(t)$
for a given ${\cal F}_l$. The results with ${\cal F}_l$
from (\ref{DirFactSimple}), (\ref{DirFactProxy}), and (\ref{dpdtnonlocal})
are very similar to each other, so we present only
the scalp--price calculated from  (\ref{DirFactProxy}) terms; \texttt{\seqsplit{.dp\_to\_use}=F\_dpdt0\_SCALP}
in \hyperref[ScalpedMaxIProjection]{\texttt{\seqsplit{ScalpedMaxIProjection.java}}}.
The regular price is a sum of all price changes (\ref{Fleqdp}),
the scalp--price is a sum of relevant to market dynamics (high $I$)
price changes (\ref{PriceFromdF}). In Fig. \ref{PPScalpedWH}
regular and scalp--price are presented. One can clearly see,
that while the regular price has an erratic behavior
due to whatever market moves, the scalp--price ${\cal P}$
has a more regular type of behavior. If scalp--price
changes it's trend --- the trend actually changes.
The scalp--price ${\cal P}$ (\ref{PriceFromdF})
is defined within a constant, and it is typically not  a good
idea to compare regular and scalp--price.
However, if one takes an event in the past,
where the price is equal to the last price,
the change in the scalp--price gives marker direction,
i.e. instead of comparing price and scalp--price,
one needs to identify a situation of zero price change,
then scalp--price change gives market directional information.
From a market practitioner's
perspective, plain observation of the scalp--price
is a good source of directional information.
As we have discussed above,
a typical  price behavior after liquidity excess (high $I$)  event 
is to bounce a little, then go in the original direction of the market.
This gives a risk of on what to bet: ``little bounce'' or ``follow the market''. The ${\cal P}$, obtained from (\ref{DirFactProxy}) ${\cal F}_l$,
has no ``little bounce''
contributions; watching the ${\cal P}$ is actually
watching pure market trend.
If the price moves, and the scalp--price stays --- this typically
indicates a bear market rally or a bull market sell--off.
The ${\cal P}$ is an integral attribute. The ${\cal F}=d{\cal P}/dt$
is a local attribute. One can try the
\begin{align}
  \Braket{\psi_I^{[IH]}|{\cal F}|\psi_I^{[IH]}}&=
  \Braket{\psi_I^{[IH]}|\frac{d{\cal P}}{dt}|\psi_I^{[IH]}}
  \label{localattribdPcal}
\end{align}
attribute (not show in Fig. \ref{PPScalpedWH},
see \texttt{\seqsplit{.F\_IH}} field
of \hyperref[ScalpedMaxIProjection]{\texttt{\seqsplit{ScalpedMaxIProjection.java}}} output), but the result is worse compared to the ${\cal P}$ result,
no clear bear/bull market switch is observed.
The situation is similar to the one in Fig. \ref*{Mu-figdpdtdirect} of
Ref. \cite{ArxivMalyshkinMuse}: $dp/dt$ chart in the $\Ket{\psi_I^{[IH]}}$ state.

\subsection{A Demonstration of the Directional Information}
\begin{figure}[p]
  %
  \includegraphics[width=16cm]{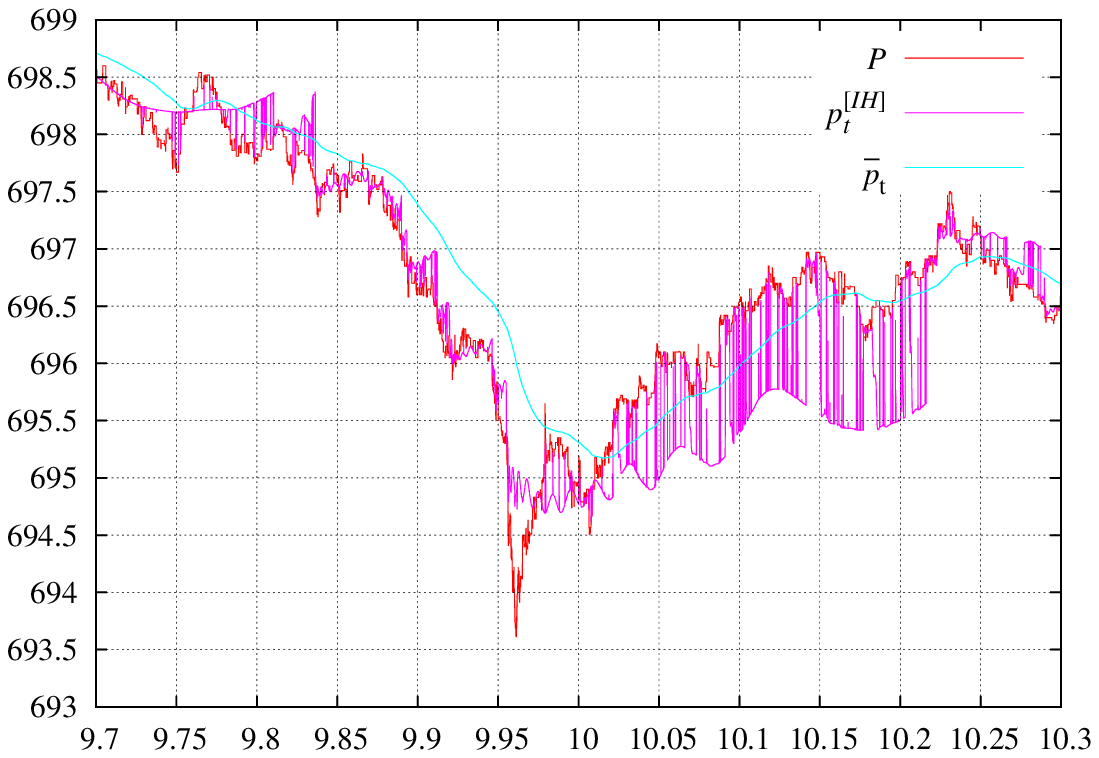}
  \includegraphics[width=16cm]{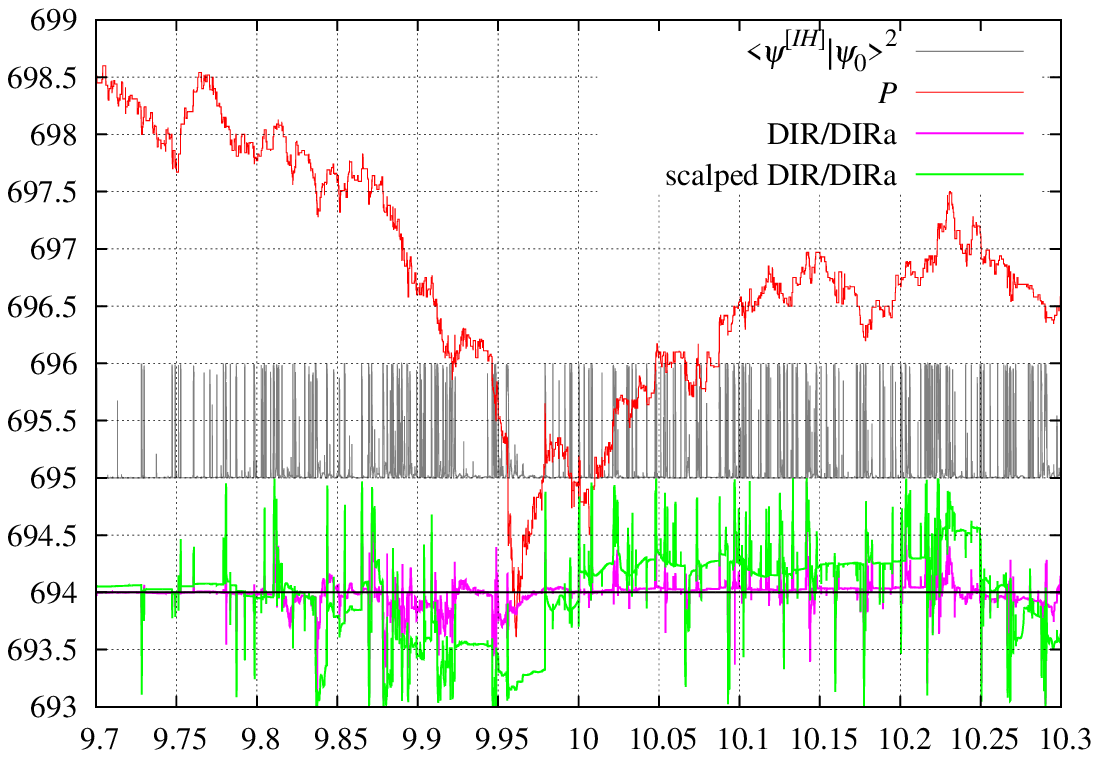}
  \caption{\label{DirAnswer}
    Top: $\frac{\Braket{p}}{\Braket{1}}$ (moving average),
    and $p_t^{[IH]}$ (\ref{ptDefine}).
    Bottom:The $\mathrm{DIR}=P^{last}-\Braket{\psi_I^{[IH]}|p|\psi_I^{[IH]}}$ (\ref{pLdiff})
    and $\mathrm{DIR\_scalped}={\cal P}^{last}-\Braket{\psi_I^{[IH]}|{\cal P}|\psi_I^{[IH]}}$ (\ref{pLdiffWSum}); both DIRs are
    \textbf{normalized} to all ${\cal F}_l$ taken positive (normalized to total variation). }
\end{figure}
The directional information should be accumulated
over an interval of a substantial duration
for the reason of low information available in a single price change.
However, the strategies as the last price minus the average
will never work for the reasons of fixed time scale
of price averaging.
In \cite{2015arXiv151005510G}, the time--scale
of the state of maximal past $I$, the  $\Ket{\psi_I^{[IH]}}$,
was introduced and the (\ref{pLdiff}) answer was obtained.
In this paper, the next critically--important step is made:
instead of regular price $p$, the scalp--price ${\cal P}$ (it includes
only high $I$ events: only  relevant to market dynamics price moves)
is introduced and the (\ref{pLdiffWSum}) answer is obtained.
In the Fig. \ref{DirAnswer} (bottom) two directional answers are presented.
In the top chart moving average $\frac{\Braket{p}}{\Braket{1}}$
and $p_t^{[IH]}$ are presented.
In the bottom chart the
(\ref{pLdiff}) (\texttt{\seqsplit{.dp\_to\_use=F\_SAMPLE\_DP\_NOSCALP}}),
and
(\ref{pLdiffWSum}) (\texttt{\seqsplit{.dp\_to\_use=F\_dpdt0\_SCALP}}),
they are normalized to the same integral taken with all ${\cal F}_l$
positive in (\ref{pLdiff}) and
(\ref{pLdiffWSum}).
One can clearly see that:
\begin{itemize}
\item When divided by the
  absolute variation,
  the non---scalped answer (\ref{pLdiff})  is pretty small,
  and the scalped one (\ref{pLdiffWSum}) is much larger.
  This means that the price can be moved due to a variety
  of reasons, and only scalped price changes (\ref{FdP})
  are relevant to the market dynamics. Moreover, high $I$
  market moves are much more consistent.
\item
  Look at $t\in[9.9\dots 9.95]h$ interval.
  The price bounce around $p_t^{[IH]}$,
  what make it difficult to trade the direction as $P^{last}-p_t^{[IH]}$.
  In the same time the scalp--price (\ref{pLdiffWSum}) stays in the
  same sign, the scalp function ${\cal S}(t)$ (\ref{WprojIH})
  is about zero in this interval, see Fig. \ref{PPScalpedWH}
\item
  Look around $t=10h$. The scalped answers captured
  all the relevant price changes and switched from bear to bull market.
  The execution flow $I$ defines market sentiment.
\end{itemize}

\subsection{On the Selection of ${\cal F}_l$ }
\begin{figure}[p]
  \includegraphics[width=16cm]{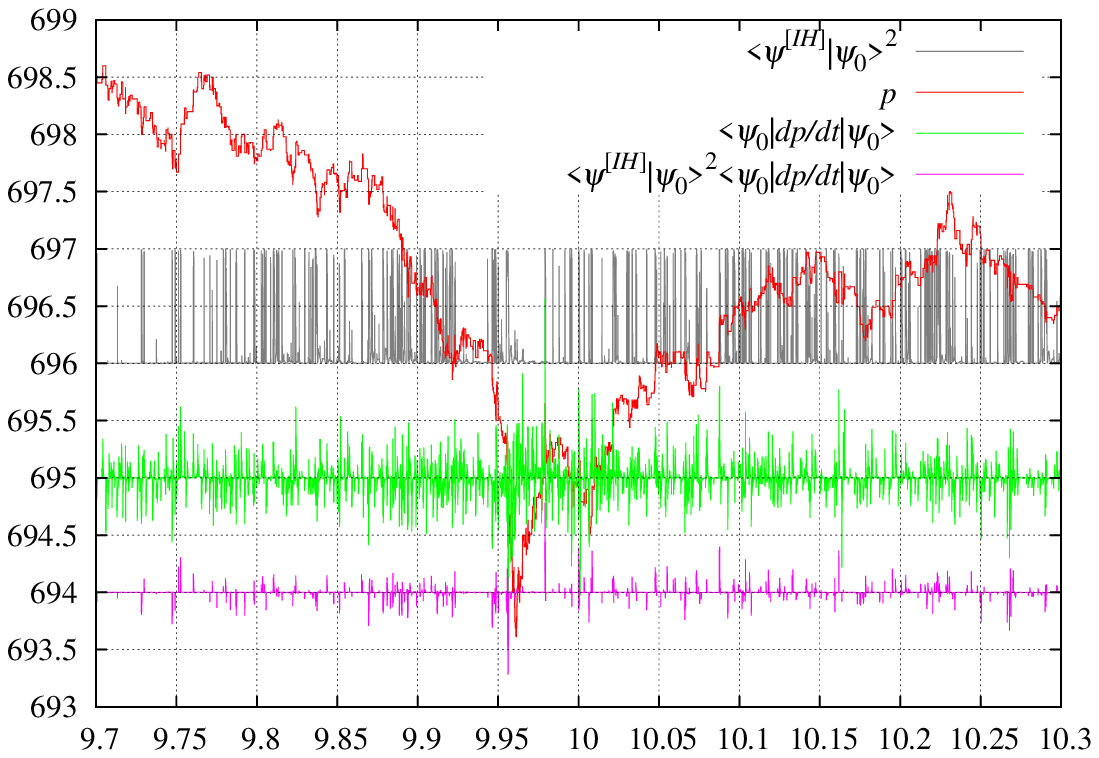}
  \includegraphics[width=16cm]{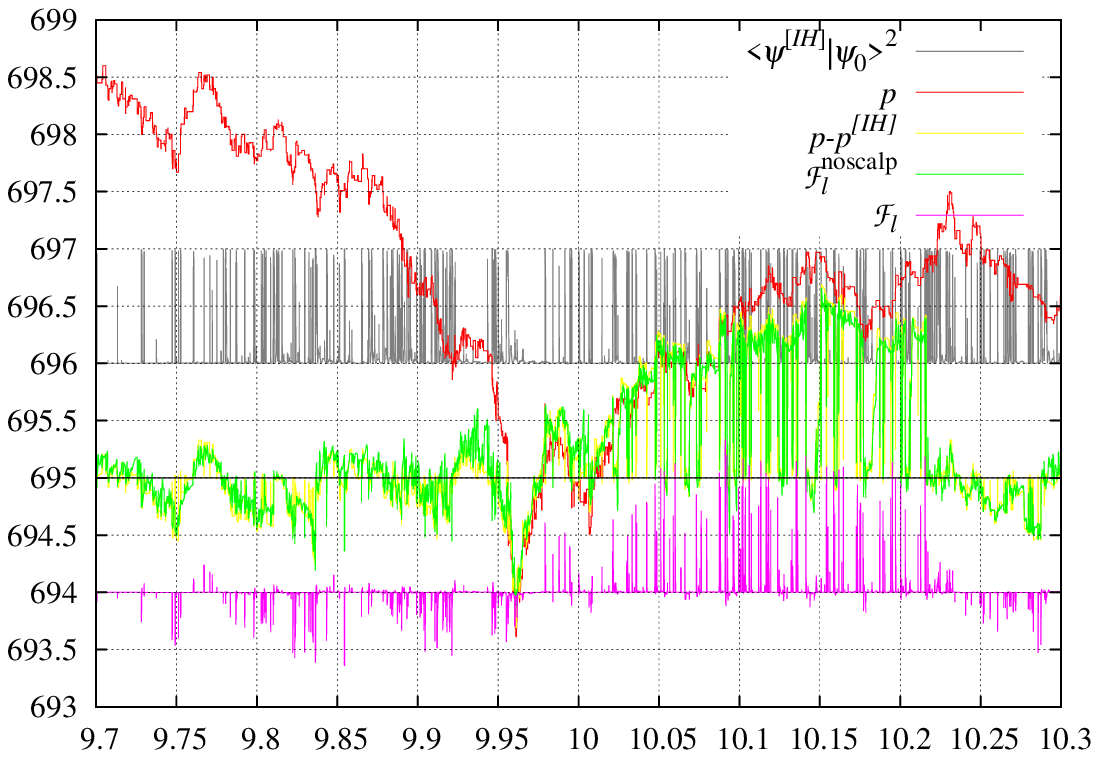}
  \caption{\label{FlSelectionFig}
    Top: the scalp--function $\cal S$ (\ref{WprojIH}),
    $\Braket{\psi_0|dp/dt|\psi_0}$, and ${\cal F}_l$ from (\ref{DirFactProxy}),
    Bottom: same $\cal S$ (\ref{WprojIH}),
    ${\cal F}_l$ from (\ref{FlProbabilityCorrelation2DirSignedScalp})
    for $z=\lambda_{p^*}^{[1]}- \lambda_{p^*}^{[0]}$
    with (pink) and without (green) scalp function multiplied;
    the $p-p^{[IH]}$ (yellow) is also presented.    
    The values are shifted to 694, 695, and 696 levels and scaled
    to fit the chart.
}
\end{figure}
The selection of ${\cal F}_l$ to be summed to the scalp price ${\cal P}$
(\ref{PriceFromdF})
is the most important question for directional attribute selection.
Consider two choices. In Fig. \ref{FlSelectionFig} top
${\cal F}_l=\Braket{\psi_0|dp/dt|\psi_0}$ from (\ref{DirFactProxy})
is presented. One can see that $dp/dt$ (green) has rather erratic behavior,
that is
caused by a variety of market moves, the sum of these moves gives
the regular price $p$. But when each price move is multiplied by
the scalp--function
${\cal S}=\Braket{\psi_I^{[IH]}|\psi_0}^2$ from (\ref{WprojIH}),
this selects only high $I$ market moves,
what makes the directional behavior much more clear (pink),
the sum now gives the scalp price $\cal P$
from the Fig. \ref{PPScalpedWH}.
But even in this simplistic case
the scalp--price selects only ``relevant'' market moves.

A much more interesting behavior can be observed with
${\cal F}_l$ from (\ref{FlProbabilityCorrelation2DirSignedScalp}).
Selecting $\Ket{\psi_0}$, $\Ket{\psi_I^{[IH]}}$ basis,
solving (\ref{lamevP}), then obtaining (\ref{FlProbabilityCorrelation2DirSignedScalp})  --- this
 accumulates much more directional
information in ${\cal F}_l$.
When $\Ket{\psi_I^{[IH]}}$ and $\Ket{\psi_0}$
are not close to each other,
the (\ref{FlProbabilityCorrelation2DirSignedScalp})
for $z=\lambda_{p^*}^{[1]}- \lambda_{p^*}^{[0]}$
is approximately equals to last price and $p^{[IH]}$
difference multiplied by the scalp--function (that is close to zero).
When $\Ket{\psi_I^{[IH]}}$ and $\Ket{\psi_0}$
are close to each other
the (\ref{FlProbabilityCorrelation2DirSignedScalp})
does not vanish\footnote{In this case
 ${\cal F}_l$ from (\ref{DirP0PiH})
is almost zero, but ${\cal F}_l$  from (\ref{DirPiPtroxy})
does not vanish, while providing
a much smaller response than (\ref{FlProbabilityCorrelation2DirSignedScalp})},
i.e. the (\ref{FlProbabilityCorrelation2DirSignedScalp})
does not vanish (like $p-p^{[IH]}$, yellow) when $I_0$ is extremely high.
However, the ${\cal F}_l$ enters into the integral (\ref{pLdiffWSum}),
and the selection of the $z$ is a non--trivial task.
The most  important feature
of the charts in Fig. \ref{FlSelectionFig}
is that once we got a spike 
in the $\Ket{\psi_I^{[IH]}}$ state ---
the trend is going to continue.
These spikes are much greater in values (because of non--local price difference)
compared to
local price difference $p(t_l)-p(t_{l-1})$ of Eq. (\ref{DirFactSimple}).
This allows to collect much more directional information,
than can can be typically obtained from price changes.

We can generalize this non--local price change approach.
Consider $p_t^{[IH]}$ in Fig. \ref{DirAnswer} ($p^{[IH]}$ is very close to it).
A typical behavior for $p^{[IH]}$ is to jump from some past value
to last price when the execution flow $I_0$ becomes large,
(\ref{WprojIH}) is the criteria of $I_0$ largeness. How often these jumps
occur is the criteria to determine market direction,
see Fig. \ref{FlSelectionFig}.
These non--local structural 
changes in $\Ket{\psi_I^{[IH]}}$
can be included to scalp--price,
for every tick $l$ calculate:
\begin{subequations}
  \label{tlcalcs}
\begin{align}
  &p^{[IH]}(t_l)&  \label{tlcalcs:pIH}\\
  &\lambda_I^{[IH]}(t_l) \label{tlcalcs:lambdaI} \\
  &{\cal S}(t_l)=\Braket{\psi_I^{[IH]}|\psi_0}^2 \label{tlcalcs:S} \\
  &\text{All (\ref{tlcalcs}) values are calculated from the sequence: $(t_m,p(t_m),V(t_m)); m=1\dots l$} \nonumber
\end{align}
\end{subequations}
These are just Eq. (\ref{GEVI}) solution
performed for every observation tick $l$
using $m=1\dots l$ previous ticks as input data.
This is the result we had obtained back in Ref. \cite{2015arXiv151005510G}.
The \textbf{new} idea is to consider the $p^{[IH]}(t_l)$ as if it were the last price $p(t_l)$.
This way one tick price change
becomes non--local:
\begin{align}
   p(t_l)-p(t_{l-1})&\to p^{[IH]}(t_l)-p^{[IH]}(t_{l-1})  \label{dptodFpIH}
\end{align}
Depending on the execution flow,
the $\Ket{\psi_I^{[IH]}}$ may (or may not)
change drastically at every tick.
One tick non--local difference  $p^{[IH]}(t_l)-p^{[IH]}(t_{l-1})$ can be much greater than one tick local price difference
$p(t_l)-p(t_{l-1})$,  see Fig. \ref{DirAnswer}. As we discussed in the introduction
only \textsl{low $I$ $\to$ high $I$} to be considered:
\begin{align}
  \theta_{I+}(t_l)&=
  \begin{cases}
   1        & \text{if}\: \lambda_I^{[IH]}(t_l)\ge\lambda_I^{[IH]}(t_{l-1}) \\
   0        & \text{otherwise}
\end{cases} \label{thetaPlus}\\
  (t_l-t_{l-1}){\cal F}_l&=
  z\left[p^{[IH]}(t_l)-p^{[IH]}(t_{l-1})\right]
  \theta_{I+}(t_l)
  \label{FlDpIH}
\end{align}
With a number of possible options for $z$:
\begin{subequations}
  \label{zselkpih}
\begin{align}
  z&=1 \label{zconst} \\
  z&= {\cal S}(t_l)\left[V(t_l)-V(t_{l-1})\right] \label{zdV} \\
  z&= dt {\cal S}(t_l) \left[\lambda_I^{[IH]}(t_l)-\lambda_I^{[IH]}(t_{l-1})\right] \label{zdI1}\\
  &\dots \nonumber
\end{align}
\end{subequations}
This is the ${\cal F}_l$ to be used in (\ref{pLdiffWSum}).
The (\ref{FlDpIH}) considers every \textsl{low $I$ $\to$ high $I$} jump
in $p^{[IH]}(t_l)$ (not in $p(t_l)$) as the source of the directional information.

\begin{figure}[ht]
  %
  \includegraphics[width=16cm]{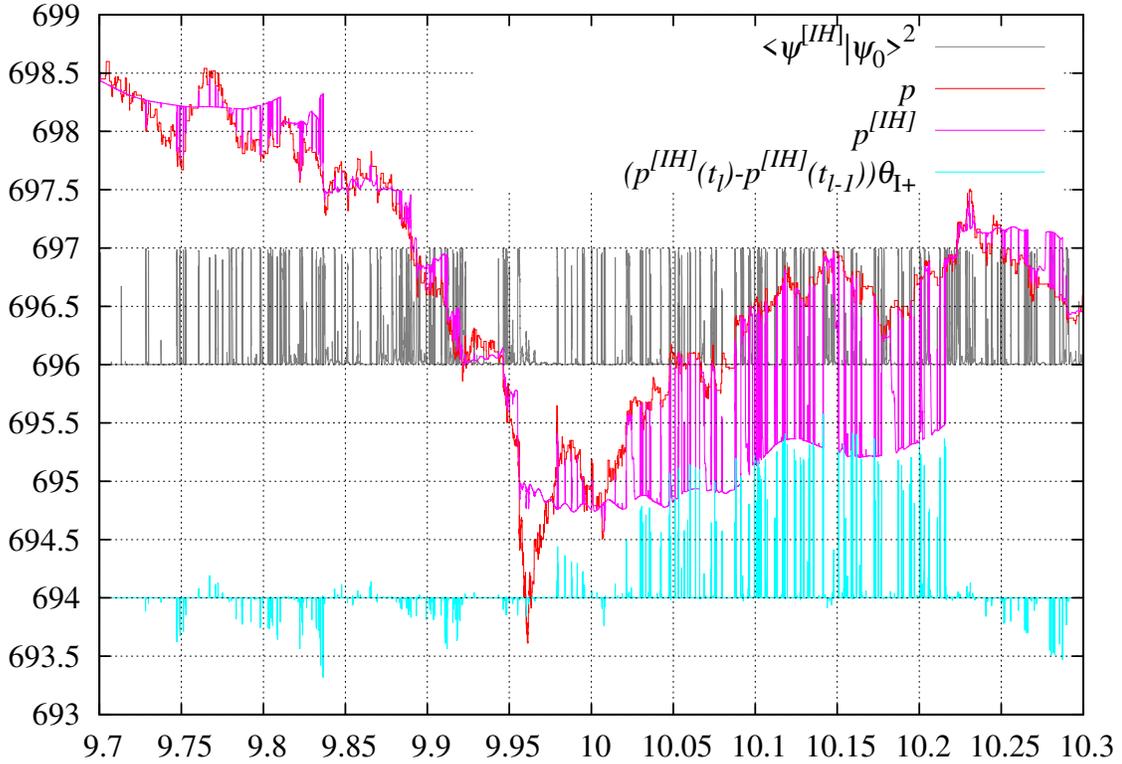}
  \caption{\label{FlPIHExampleFig}
    The price, scalp--function $\Braket{\psi_I^{[IH]}|\psi_0}^2$,
    $p^{[IH]}$ (\ref{pIH}) (pink), and ${\cal F}_l$ from (\ref{FlDpIH})
    without $z$ term (blue). ${\cal F}_l$ is presented
    as one tick
    $p^{[IH]}$ change $p^{[IH]}(t_l)-p^{[IH]}(t_{l-1})$
    multiplied by $\theta_{I+}(t_{l})$ factor (\ref{thetaPlus}).
}
\end{figure}

In Fig. \ref{FlPIHExampleFig} a demonstration of non--local
price change (\ref{dptodFpIH}) is presented. Only $p^{[IH]}(t_l)-p^{[IH]}(t_{l-1})$
with positive $\lambda_I^{[IH]}(t_l)-\lambda_I^{[IH]}(t_{l-1})$ are presented
(the Eq. (\ref{FlDpIH}) with $z=1$ and $dt=1$).
On can clearly see that
the non--local directional information is:
\begin{itemize}
\item Much greater than the local price change $p(t_{l})-p(t_{l-1})$.
\item The bull/bear market trend switch can be much better identified.
  The $p^{[IH]}(t_l)-p^{[IH]}(t_{l-1})$ with (\ref{thetaPlus}) constaint
  preserves the sign during extended intervals.
\item The ``bounce back'' interval $t\in[9.9\dots 9.95]h$
  is clearly identified: it has no $I$ spikes, the $\Ket{\psi_I^{[IH]}}$
  does not change, and the $p^{[IH]}(t_l)-p^{[IH]}(t_{l-1})$ is close
  to zero even without ${\cal S}$ multiplied!
\end{itemize}
This makes us to conclude,
that non--local price change (\ref{dptodFpIH})
taken with the constraint (\ref{thetaPlus}) provides
a very promising
possible directional indicator.
Fig. \ref{FlPIHExampleFig} presents a
non--local price answer (\ref{dptodFpIH})
obtained from one tick  $p^{[IH]}$ price change $p^{[IH]}(t_l)-p^{[IH]}(t_{l-1})$
as it were one tick regular price change $p(t_{l})-p(t_{l-1})$.
This answer is similar (but much better) than
Fig. \ref{FlSelectionFig} (bottom)
answer, that is obtained from the regular moments by solving (\ref{lamevP}) $d=2$
eigenvalue problem.
The (\ref{FlDpIH}) is the directional indicator.
However, because it enters the integral (\ref{PriceFromdF}),
the selection of proper integration
weight $z$ is required. This to be a subject of a separate
study. In the simplest form a non--local answer
can be obtained from (\ref{tlcalcs}) solution of (\ref{GEVI}) problem,
then consider:
\begin{itemize}
\item Only $\lambda_I^{[IH]}(t_l)\ge\lambda_I^{[IH]}(t_{l-1})$ events: $\theta_{I+}(t_l)>0$, Eq. (\ref{thetaPlus}), field (\ref{scalpPrice:fielddIG}).
\item For such events consider $p^{[IH]}(t_l)-p^{[IH]}(t_{l-1})$
  as it were one tick price change $p(t_l)-p(t_{l-1})$, Eq. (\ref{dptodFpIH}),
  field (\ref{scalpPrice:fielddpIG}).
  In Fig. \ref{FlPIHExampleFig} an example of such a non--local price changes
  is presented.
\end{itemize}

\section{\label{directionalInfo}On the Directional Information Calculation}
Given very interesting results of the previous section,
let us formulate all the components, required to
obtain directional information
from (time, execution price, shares traded)
market observations triples,
and how these components can be improved.
\begin{itemize}
\item The state important for market dynamics.
  The answer we have is (\ref{rhoIH}). Other states (such as considered in the Appendix
  \ref{VdivT}) can be also tried. In any case such a state is obtained 
  from regular moments (\ref{inputmoments:qm}) and (\ref{inputmoments:Iqm}),
  solving some kind of $I\xrightarrow[{\psi}]{\quad }\max$ problem.
  The solution gives us open/close position signals
  and the scale for directional calculations.
\item
  The problem to obtain the direction
  is way more complex, it requires scalp--moments (\ref{pLdiffWSum}).
  For the scalp--function ${\cal S}$ the best\cite{malyshkin2018spikes}
  answer is (\ref{WprojIH}). For ${\cal F}_l$
  several answers (\ref{DirFactSimple}), (\ref{DirFactProxy}),
  (\ref{DirPiPtroxy}), and (\ref{dpdtnonlocal}) produce good results,
  that are very similar to each other,
  the non--local answer (\ref{FlDpIH}) is of special interest.
  The ``varied'' answers (\ref{DirpIProxy}), and (\ref{DirpIProxydPi})
  are worse with and without scalp--function
  multiplied.
  The simplest practical abswer is the 
  scalp--moments directional answer (\ref{pLdiffWSum}),
  as a scale one can use absolute variation: take all ${\cal F}_l$
  positive in (\ref{pLdiffWSum}).
  However, a number of non--local answers of (\ref{FlDpIH})
  type can be obtained utilizing (\ref{tlcalcs}) and (\ref{dptodFpIH}).
\end{itemize}

\section{\label{concl}Speculations}

The scalp--moments are price change moments \textsl{filtered}
by high $I$ events: $I$ is the driving force of the market.
The question arises whether a directional information
can be obtained from the regular moments (\ref{inputmoments})?
We are inclined to say no.
A number of constrained (see Appendices \ref{PnLRatio} and \ref{ConstrainsOperator} below)
and unconstrained optimization
problems have been tried (among many others)
 without any success at obtaining market
directional information:
\begin{subequations}
  \label{Tried:withoutSuccess}
\begin{align}
\max\limits_{\psi}&\frac{\Braket{\psi|(p-p^{[IH]})^2I|\psi}}
  {\Braket{\psi|\psi}}
  & &\text{\texttt{DynHPnL.java}}
  \label{pIHPnL} \\
\max\limits_{\psi}&\frac{\Braket{\psi|(p-P^{last})^2I|\psi}}
  {\Braket{\psi|\psi}}
  & &\text{\texttt{DynPnL.java}, \texttt{PnLSensitivity.java}}
  \label{plastPnL} \\
\max\limits_{\psi}&\min\limits_{p_x}\frac{\Braket{\psi|(p-p_x)^2I|\psi}}
  {\Braket{\psi|\psi}}
  & &\text{\texttt{DynYp.java}, \texttt{DIminP2maxI.java}}
  \label{p2Imin} \\
\max\limits_{\psi}&\max\limits_{p_x}\frac{\Braket{\psi|I|\psi}}
  {\Braket{\psi|(p-p_x)^2|\psi}}
  & &\text{\texttt{DynYp.java}, \texttt{DIminP2tmaxI.java}}
  \label{p2mint} \\      
\max\limits_{\psi}&\min\limits_{p_1,p_2}\frac{\Braket{\psi|(p-p_1)^2(p-p_2)^2I|\psi}}{\Braket{\psi|\psi}}
  & &\text{Ref. \cite{ArxivMalyshkinMuse} Section \ref*{Mu-PsiFutureIIter}, \texttt{PnLdIV4.java}} \label{P4Imin} \\             
\max\limits_{\psi}& \frac{\Braket{\psi|\left(p-p_t\right)^2I|\psi}}
                         {\Braket{\psi|\psi}}
& & \text{\texttt{MaxPtPv2I.java}, \texttt{MinMaxPnLratioNorm.java}}
                   \label{Pt2Nom} \\ 
\max\limits_{\psi}&\frac{\Braket{\psi|\left(p-p_v\right)^2I|\psi}}
                   {\Braket{\psi|\left(p-p_t\right)^2|\psi}}
  & &\text{\texttt{MaxPnLratio.java} ; \texttt{flag\_swap\_PtPv=false}}
                   \label{PvPtNotSwapped} \\
\max\limits_{\psi}&\frac{\Braket{\psi|\left(p-p_t\right)^2I|\psi}}
                   {\Braket{\psi|\left(p-p_v\right)^2|\psi}}
  & &\text{\texttt{MaxPnLratio.java} ; \texttt{flag\_swap\_PtPv=true}}
                   \label{PvPtSwapped} \\
  \max\limits_{\psi}&\frac{\Braket{\psi|I|I|\psi}}
  {\Braket{\psi|(p-P^{last})^2I|\psi}}
  & & \text{\texttt{MaxPPl2I.java}, \texttt{MaxPPl2Iinbasis.java}}
    \label{MaxPPl2}
\end{align}
\end{subequations}
The regular moments answers are: 1. not ``sufficiently sharp'',
see Appendix \ref{psiX}, and 2. price changes sum is small
relatively total variation, see Fig. \ref{DirAnswer}.
In the same time, when we go to the scalp--moments  (\ref{DirFact})
these problems get solved.

When, in September 1997, I joined Columbus Advisors LLC (Greenwich CT),
the fund had been doing Emerging Market sovereign fixed income
convergence--divergence relative value spread trades.
The following year, I studied a classic technical analysis
book with the goal to program some of the rules algorithmically.
However, I was not able to program even \textsl{a single rule} from the book.
The reason was simple: any rule required a time scale to apply.
Time scale selection
is the main criterion separating good traders from bad, and the criterion which defines a trader's talent.
The state (\ref{rhoIH}) is an algorithmic criterion,
that automatically determines the time scale.
This  criterion is actually very simple ideologically:
look back to find an event of trading with maximal $I$.
 The time between this event
and ``now''
is the time scale.
The typical market practitioner's activity is to watch the  difference
between the last price and moving average calculated
on the time scale obtained his feel.
With a proper time scale,
any strategy (like return to the moving average) would work,
and Ref. \cite{2015arXiv151005510G} answer of last price minus $p^{[IH]}$ (\ref{pIH})
was my first successful attempt.

Besides the time scale, the most important result of this paper is
that ``not all price moves are equal''. We need to select
only the high $I$ price moves\footnote{
  I think that the market impact concept is a dead end.
  }.
High execution rate requirement
is the condition creating an asymmetry to separate 
the
``bounce a little, then to go in the original direction of the market''
and ``go in the original direction of the market straight away''
scenarios, such as to identify a
\href{https://en.wikipedia.org/wiki/Dead_cat_bounce}{bear market rally}
on steroids.
The answer we obtained is the scalp--price (\ref{PriceFromdF}).
It does not have any ``internal averaging'',
but in the same time it has all low $I$ price changes removed!
This way, the scalp--price has no ``bounce a little'' behavior.
Only hardcore. Only directional. See the Fig. \ref{PPScalpedWH}.
\href{http://www.ioffe.ru/LNEPS/malyshkin/AMuseOfCashFlowAndLiquidityDeficit.zip}{The software} is available\cite{polynomialcode} under the \href{https://www.gnu.org/licenses/gpl-3.0.en.html}{GPLv3} license.

\begin{acknowledgments}  
  Vladislav Malyshkin would like to thank
  \href{https://www.researchgate.net/profile/Alexander_Bobyl2}{Aleksandr Bobyl} from the
  \href{http://www.ioffe.ru/}{Ioffe Institute} for many discussions of
  whether the activity of a typical market participant can be reduced
  to some ``simplified automata'' with intelligence as an aggregate function of past transactions,
  and,
  even more importantly, the ability (or inability) to apply 
  the obtained knowledge to market activity; 
  \href{https://www.nytimes.com/1994/06/12/style/weddings-cathleen-ess-and-emilio-lamar.html}{Emilio J. Lamar} 
  for discussions of slow/fast
  market transitions --- especially the spread--widening effect during fast markets; and 
  \href{http://www.ioffe.ru/LNEPS/research/theory.html}{Aleksandr Maslov}
  from the \href{http://www.ioffe.ru/}{Ioffe Institute}
  for discussions on contradistinction of operators $\|dV/dt\|$ and $\|V/t\|$.
\end{acknowledgments}

\appendix

\section{\label{psiX}A demonstration of the difference between
  time and volume weighted price.}
\begin{figure}[t]
  \includegraphics[width=16cm]{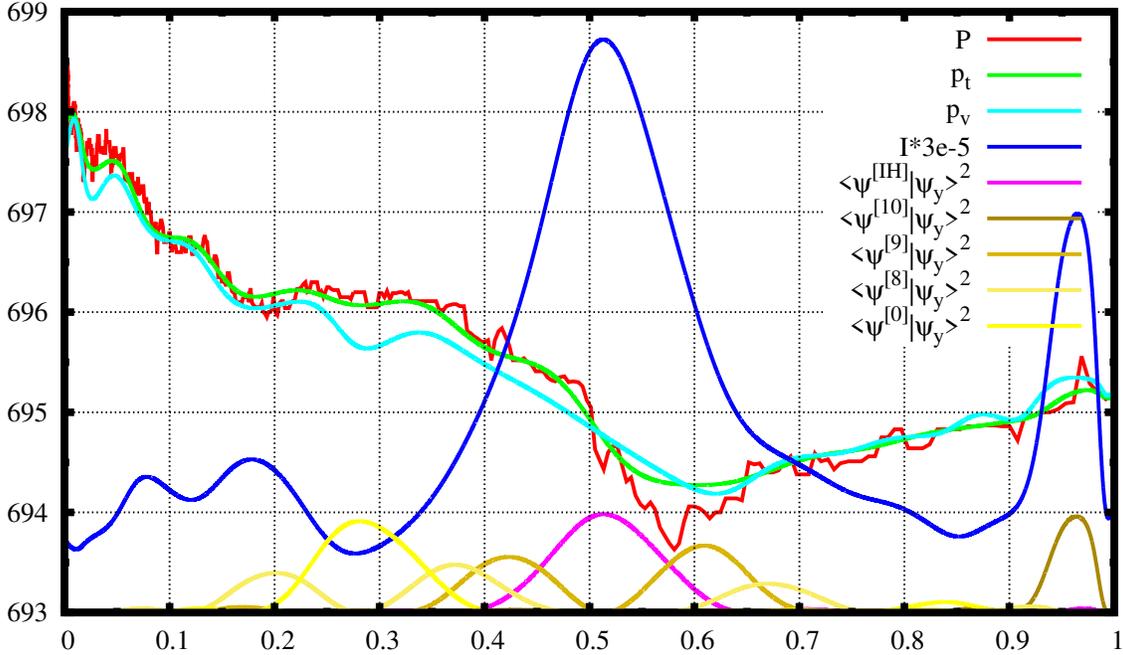}
  \caption{\label{psiback}
    The AAPL stock price on September, 20, 2012.
    Interpolation answers are calculated in Shifted Legendre basis with $n=12$ and $\tau$=128sec,
  for $0\le t \le t_{now}=9.98045$ hrs,
  $y=\exp\left((t-t_{now})/\tau\right)$, $y=[0\dots 1]$. 
  Execution flow (\ref{Iy}), time (\ref{Pty}), and volume (\ref{Piy}) weighted prices are presented.
  One can clearly see the $p_v(y)-p_t(y)$ changes the sight at $y$,
  corresponding to a high $I$.
  The maximal eigenstate $IH$, (\#11=$n-1$), 
  $\Braket{\psi_y(x)|\psi_I^{[IH]}}^2$, pink, is typically a localized state.
  The projections $\Braket{\psi_y(x)|\psi_I^{[i]}}^2$
  on four other eigenstates (\#0, \#8, \#9, and \#10), yellow, are presented
  as an example of delocalized states.  
  The execution flow $I$ and the projection
  are shifted to $693$ to fit the chart.
  }
\end{figure}
To demonstrate the difference consider localized at $x=y$ the wavefunction
$\psi_y(x)$ (\ref{psiRN}),
producing 
Radon--Nikodym interpolating answer,
 Eq. (7) of Ref. \cite{2016arXiv161107386V},
Different attributes (price, execution flow, etc.)
are interpolated using the $\psi^2_y(x)\omega(x)dx$ weight:
\begin{eqnarray}
  \psi_y(x)&=&\frac{\sum\limits_{j,k=0}^{n-1}Q_j(x)G^{-1}_{jk}Q_k(y)}
  {\sqrt{\sum\limits_{j,k=0}^{n-1}Q_j(y)G^{-1}_{jk}Q_k(y)}}
  \label{psiRN}\\
  1&=&\Braket{\psi_y|\psi_y} \\
  I(y)&=&\Braket{\psi_y|I|\psi_y}\Big/\Braket{\psi_y|\psi_y} \label{Iy}\\
  p_t(y)&=&\Braket{\psi_y|p|\psi_y} \Big/\Braket{\psi_y|\psi_y} \label{Pty} \\
  p_v(y)&=&\Braket{\psi_y|pI|\psi_y}\Big/\Braket{\psi_y|I|\psi_y} \label{Piy}
\end{eqnarray}
One can see that:
\begin{itemize}
\item For a large $n$ (we use $n=12$) the $p_t$ and $p_v$ are very similar.
\item The projection $\Braket{\psi_y(x)|\psi_I^{[IH]}}^2$ is close
  to $1$ for large $I$, i.e. the $\psi_I^{[IH]}(x)$ is typically
  a localized function,
  this is not the case for other states. See four other eigenstates projections (yellow).
\item The $p_v-p_t$ changes the sign at large  $I$.
  Only the states with a large $dI/dt$ provide
   weight asymmetry required to obtain directional information
  using $dV$ vs. $dt$ averaging. 
\end{itemize}

\section{\label{codestructure}Computer Implementation}
The codebase architecture is described in the Appendix  \ref*{Mu-codestr}
of Ref. \cite{ArxivMalyshkinMuse}.
Relevant to this paper functionality consists of:
\begin{itemize}
  \item
  Conversion of the transaction sequence of an observable $f$
  to a vector of moments $\Braket{fQ_m}$, $m=[0\dots 2n-2]$,
  several bases $Q_k(x)$ are implemented ($x=t$,  $x=\exp\left(-(t_{now}-t)/{\tau}\right)$, and $x=p(t)$, see the Section \ref*{Mu-basis} of Ref. \cite{ArxivMalyshkinMuse}), the integration
  measure is always exponential decay:
  $d\mu=\exp\left(-(t_{now}-t)/{\tau}\right)dt$.
  See the classes \texttt{\seqsplit{com/polytechnik/trading/\{QVMDataL,QVMDataP,QVMData\}.java}}
\item
  Using basis functions multiplication operator (Eq. (\ref*{Mu-cmul}) of Ref. \cite{ArxivMalyshkinMuse}),
  obtain the $\Braket{Q_j|f|Q_k}$, $j,k=[0\dots n-1]$ matrix
  from the moments $\Braket{fQ_m}$, $m=[0\dots 2n-2]$.
\item
  There are a number of observables $f$ possibly to consider (price, price change, execution flow, etc.). Depending on the approach used,
  a different set of observables is required. All the $\Braket{Q_j|f|Q_k}$
  matrices we possibly use in this paper are stored in the class
  \texttt{\seqsplit{com/polytechnik/trading/SMomentsData.java}}.
\item
  If/When, in addition to a $\Braket{Q_j|f|Q_k}$ matrix,
  the matrix corresponding to the derivative $df/dt$ (or to the integral $\int^{t}f(t^{\prime})dt^{\prime}$), is required,
  then,
  for a basis with infinitesimal time--shift operator $D(Q(x))$,
  the result can be obtained using integration by parts, see Appendices \ref{VMatrixElements} and \ref{dIdtMatrixElements}.
\end{itemize}
As a result of these preliminary steps
the  $n\times n$ matrices are obtained:
$\Braket{Q_j|Q_k}$, $\Braket{Q_j|\frac{dp}{dt}|Q_k}$, $\Braket{Q_j|p|Q_k}$, $\Braket{Q_j|p^2|Q_k}$, $\Braket{Q_j|p^3|Q_k}$,
$\Braket{Q_j|I|Q_k}$,
$\Braket{Q_j|pI|Q_k}$, $\Braket{Q_j|p^2I|Q_k}$, and $\Braket{Q_j|p^3I|Q_k}$.
These are plain
\href{https://en.wikipedia.org/wiki/Moving_average#Exponential_moving_average}{exponential moving--average}
of: an observable $f$ multiplied by
two basis functions product $Q_j(x)Q_k(x)$;
for example if $f=p$, then $\Braket{Q_0|p|Q_0}$ is 
exponential moving average of price.

\subsection{\label{EVXData}The \texttt{EVXData.java} implementation}
The class
\texttt{\seqsplit{com/polytechnik/utils/EVXData.java}}
takes two matrices $\Braket{Q_j|f|Q_k}$, $\Braket{ Q_j|Q_k}$
and basis functions operations class (extending the
\texttt{\seqsplit{com/polytechnik/utils/OrthogonalPolynomialsABasis.java}}),
solves generalized eigenvalue problem, such as (\ref{GEVI}) for $\|f\|=\|I\|$,
and stores the result. The fields are:
\begin{subequations}
  \label{EVXData:I:fields}
  \begin{align}
    \mathtt{.sL}&=\lambda_{I}^{[IL]} \\
    \mathtt{.sH}&=\lambda_{I}^{[IH]} \\
    \mathtt{.s0}&=\Braket{\psi_0|I|\psi_0} \\
    \mathtt{.wL}&=\Braket{\psi_{I}^{[IL]}|\psi_0} \label{wLprojection} \\
    \mathtt{.wH}&=\Braket{\psi_{I}^{[IH]}|\psi_0} \label{wHprojection}
  \end{align}
\end{subequations}      
The squares $\mathtt{.wL}^2$ and $\mathtt{.wH}^2$
  are bounded to $[0:1]$, and  
  are very good indicators of whether the $I$ ``now'', the $I_0=\Braket{\psi_0|I|\psi_0}$,
  is large or small. Alternative estimator as the number of the eigenvalues
  above the $I_0$ can also be used\cite{malyshkin2018spikes}.  
  The key concept of liquidity deficit trading\cite{2015arXiv151005510G,2016arXiv160305313G}
  is to open a position at low $I_0$, large $\Braket{\psi_{I}^{[IL]}|\psi_0}^2$,
  then to close already opened position at high $I_0$, large $\Braket{\psi_{I}^{[IH]}|\psi_0}^2$, the $\mathtt{.wL}^2$ and $\mathtt{.wH}^2$
  are the indicators of these actions. The question is: whether
  to open a \textsl{long} or a \textsl{short} position
  at high $\mathtt{.wL}^2$?

\subsection{\label{ScalpedMaxIProjection}The \texttt{ScalpedMaxIProjection.java} implementation}
The class
\texttt{\seqsplit{com/polytechnik/trading/ScalpedMaxIProjection.java}}
converts a transaction sequences to a set of $\Braket{fQ_m}$ vectors,
then to a set of $\Braket{Q_j|f|Q_k}$ matrices,
stored in the object of 
\texttt{\seqsplit{com/polytechnik/trading/SMomentsData.java}} type.
Then it calls the
\hyperref[EVXData]{\texttt{\seqsplit{com/polytechnik/utils/EVXData.java}}}
class
that solves (\ref{GEVI}) eigenvalue problem.
Having the $\Ket{\psi_I^{[IH]}}$ and  $\Ket{\psi_0}$
states the scalp--function (\ref{WprojIH})
and the 
${\cal F}_l$ are obtained. Which one ${\cal F}_l$ to be used depends
on the parameter \texttt{\seqsplit{.dp\_to\_use}}.
The values \texttt{\seqsplit{F\_SAMPLE\_DP\_NOSCALP,F\_SAMPLE\_DP\_SCALP,F\_dpdt0\_SCALP,F\_varpIH\_0\_divI\_SCALP,F\_SKEWNESS\_at\_Pl\_SCALP,F\_PROBABILITYCORRELATION\_SCALP}} correspond to
(\ref{Fleqdp}), (\ref{DirFactSimple}), (\ref{DirFactProxy}),
(\ref{DirVarPsi0AZavor1e6}), (\ref{FlSkewnessPld2}), and (\ref{FlProbabilityCorrelation2DirSignedScalp})
respectively; there are several other options for \texttt{\seqsplit{.dp\_to\_use}}.
The class \texttt{\seqsplit{ScalpedMaxIProjection}}
is assumed to be called on every tick,
and the internal state is preserved in the object of
\texttt{\seqsplit{com/polytechnik/trading/StateWIScalpMomentsSaver.java}}  class.
The internal state contains an object of
\texttt{\seqsplit{com/polytechnik/trading/WIntegrator.java}} type,
that calculates the moments of the observable ${\cal F}_l$
(recurrent shift of the basis offset ($t_{now}$) for previously
calculated moments
allows the calculations to be performed extremely fast).
The \texttt{\seqsplit{WIntegrator}}  is called on every tick
with the $\texttt{\seqsplit{.Fdt}}=\left(t_l-t_{l-1}\right){\cal F}_l$
(the choice of the ${\cal F}_l$ depends on the \texttt{\seqsplit{.dp\_to\_use}} value)
to accumulate scalped data.
 The scalp--moments
are obtained by taking the \texttt{\seqsplit{.Fdt}}
instead of the $p(t_l)-p(t_{l-1})$ when calculating the (\ref{dpdtAllW}) sum.
The directional information is then obtained as (\ref{pLdiffW}).
The fields are:
\begin{subequations}
  \label{ScalpedMaxIProjection:fields}
  \begin{align}
    &\mathtt{.p\_offset} & & \text{Price offset. All prices are relatively this offset} \\ 
    &\mathtt{.pi\_average} & &\text{Volume--weighted price exponential moving average $\frac{\Braket{pI}}{\Braket{I}}$} \\
    &\mathtt{.pt\_average} & &\text{Time--weighted price exponential moving average $\frac{\Braket{p}}{\Braket{1}}$} \\
    &\mathtt{.I}& &  \text{An object of \hyperref[EVXData]{\texttt{\seqsplit{EVXData.java}}} type, (\ref{GEVI}) solution} \label{liquiditydeficitInScalp}\\
    &\mathtt{.p\_0} & &  \text{The $\Braket{\psi_0|pI|\psi_0}\big/\Braket{\psi_0|I|\psi_0}$} \label{p0field} \\
    &\mathtt{.pt\_0} & &  \text{The $\Braket{\psi_0|p|\psi_0}\big/\Braket{\psi_0|\psi_0}$} \label{p0tfield} \\
    &\mathtt{.dpdt\_0} & & \text{The $\Braket{\psi_0|\frac{dp}{dt}|\psi_0}$} \label{dpdt0field}\\  
    &\mathtt{.p\_IH}& &  \text{The (\ref{pvDefine}) in the $\Ket{\psi_I^{[IH]}}$ state (\ref{rhoIH})} \\
    &\mathtt{.pt\_IH}& &  \text{The (\ref{ptDefine}) in the $\Ket{\psi_I^{[IH]}}$ state (\ref{rhoIH})} \\
    &\mathtt{.pV\_IH}& &  \text{The (\ref{pVDefine}) in the $\Ket{\psi_I^{[IH]}}$ state (\ref{rhoIH})} \\
    &\mathtt{.pT\_IH}& &  \text{The (\ref{pTDefine}) in the $\Ket{\psi_I^{[IH]}}$ state (\ref{rhoIH})} \\
    &\mathtt{.var1pI\_IH} & & \text{The (\ref{DirpIProxy})} \label{varpiDIH} \\
    &\mathtt{.var1pI\_IH\_00} & & \text{The (\ref{DirVarPsi0AZavor1e6})} \label{varpi00} \\
    &\mathtt{pmin\_0\_IH,pmax\_0\_IH} & & \text{The eigenvalues $\lambda_{p^*}^{[0,1]}$ of (\ref{lamevP})} \label{p0IHevminmax} \\
    &\mathtt{Skewness\_0\_IH} & & \text{The ``skewness'' (\ref{SkewnessPld2})} \label {SkewnessPld2:field}\\
    &\mathtt{ProbabilityCorrelation\_0\_IH} & & \text{Directional factor $\frac{\left[\phi^{[1]}(x_0)\right]^2-\left[\phi^{[0]}(x_0)\right]^2}
      {\left[\phi^{[1]}(x_0)\right]^2+\left[\phi^{[0]}(x_0)\right]^2}$
      (\ref{ProbabilityCorrelation2Dp})} \label{FlProbabilityCorrelation2DirSignedScalp:field} \\
    &\mathtt{.I.wH} & & \text{When squared $\mathtt{.I.wH}^2$ gives the scalp function (\ref{WprojIH})} \label{scalpFunctionField} \\
    &\mathtt{.getFlFromRegularMoments()} & & \text{${\cal F}_l$ when it is from
      the moments, \texttt{NaN} otherwise}
    \label{getFlFromRegularMoments:field} \\
    &\mathtt{.sst.getSumFdt()} & & \text{The scalp--price (\ref{PriceFromdF}) with an arbitrary offset} \label{scalpPrice:field} \\
    &\mathtt{.dIH} & & \text{The $\lambda_I^{[IH]}(t_l)-\lambda_I^{[IH]}(t_{l-1})$
      difference} \label{scalpPrice:fielddIG} \\
    &\mathtt{.dp\_IH} & & \text{The $p^{[IH]}(t_l)-p^{[IH]}(t_{l-1})$
      difference} \label{scalpPrice:fielddpIG} \\
    &\mathtt{.DIR} & & \text{The (\ref{pLdiffW})} \label{DIRscalped}\\
    &\mathtt{.aDIR} & & \text{The (\ref{pLdiffW}) with all ${\cal F}_l$ taken positive} \label{DIRascalped}
\end{align}
\end{subequations}
The
liquidity deficit indicator (\ref{liquiditydeficitInScalp}) defines
whether to open or to close a position.
The  directional indicator  (\ref{DIRscalped}) from (\ref{pLdiffW})
defines, when opening a position,
whether to open a long or a short.

\subsection{\label{CallAMusestucture}The \texttt{CallAMuseOfCashFlowAndLiquidityDeficitWithScalp.java} implementation}
The class
\texttt{\seqsplit{com/polytechnik/algorithms/CallAMuseOfCashFlowAndLiquidityDeficitWithScalp.java}}
is ``an interface'' between transactions sequence input
(a tab--separated file),
liquidity deficit trading of the class
\hyperref[ScalpedMaxIProjection]{\texttt{\seqsplit{com/polytechnik/trading/ScalpedMaxIProjection.java}}},
and data output, saved as a tab--separated file.
The parameters are read by the class
\texttt{\seqsplit{com/polytechnik/algorithms/MuseConfig.java}}.
This is an example of how to run the code:
\begin{verbatim}
java com/polytechnik/algorithms/CallAMuseOfCashFlowAndLiquidityDeficitWithScalp \
      --musein_file=aapl.csv \
      --musein_cols=15:1:4:5 \
      --museout_file=museout.dat \
      --n=12 \
      --tau=128 \
      --measure=ScalpedMaxIProjectionLegendreShifted
\end{verbatim}
The parameters are:
\begin{itemize}
\item \texttt{\seqsplit{--musein\_file=aapl.csv}} : Specify input tab--separated file
  with (time, execution price, shares traded) triples time series.
  If the file is \href{https://www.gzip.org/}{\texttt{gzip}}--ed and has the \verb+.gz+ extension,
  then internal decompression
  is performed.
\item \texttt{\seqsplit{--musein\_cols=15:1:4:5}} : Out of total 15 columns
  in
  the specified \texttt{\seqsplit{--musein\_file=aapl.csv}} file, take the column \#1 as time (nanoseconds since midnight),
  \#4 (execution price), and \#5 (shares traded), column index is base 0.
\item \texttt{\seqsplit{--museout\_file=museout.dat}} : Output file name.
\item \texttt{\seqsplit{--n=12}} : Basis dimension. Typical values are:
  $n\in [4\dots 16]$.
  The $m\in [0\dots 2n-2]$
  moments (in $Q_m(x)$ basis) are calculated
  to obtain $n\times n$ matrices.  
\item  \texttt{\seqsplit{--tau=128}} : Exponent time (in seconds)
  for the measure used.
\item \texttt{\seqsplit{--measure=ScalpedMaxIProjectionLegendreShifted}} : 
  The measure. Possible values are:
  \texttt{\seqsplit{\{ScalpedMaxIProjectionLegendreShifted,ScalpedMaxIProjectionLaguerre,ScalpedMaxIProjectionMonomials\}}}, they
  correspond to the measures (\ref*{Mu-muslegendre}) and (\ref*{Mu-muflaguerre})
of Ref. \cite{ArxivMalyshkinMuse}.
The \texttt{\seqsplit{ScalpedMaxIProjectionLaguerre}}
and \texttt{\seqsplit{ScalpedMaxIProjectionMonomials}}
use the same measure, but different basis  $Q_k(x)=L_k(x)$, $x=-t/\tau$
and $Q_k(x)=x^k$, $x=t/\tau$ respectively.
These two results 
  should be \textsl{identical},
  as the measure is the same, and all the calculations are
  $Q_k(x)$--basis invariant (but the numerical stability can be drastically different).
\end{itemize}

\begin{figure}[t]
  \includegraphics[width=16cm]{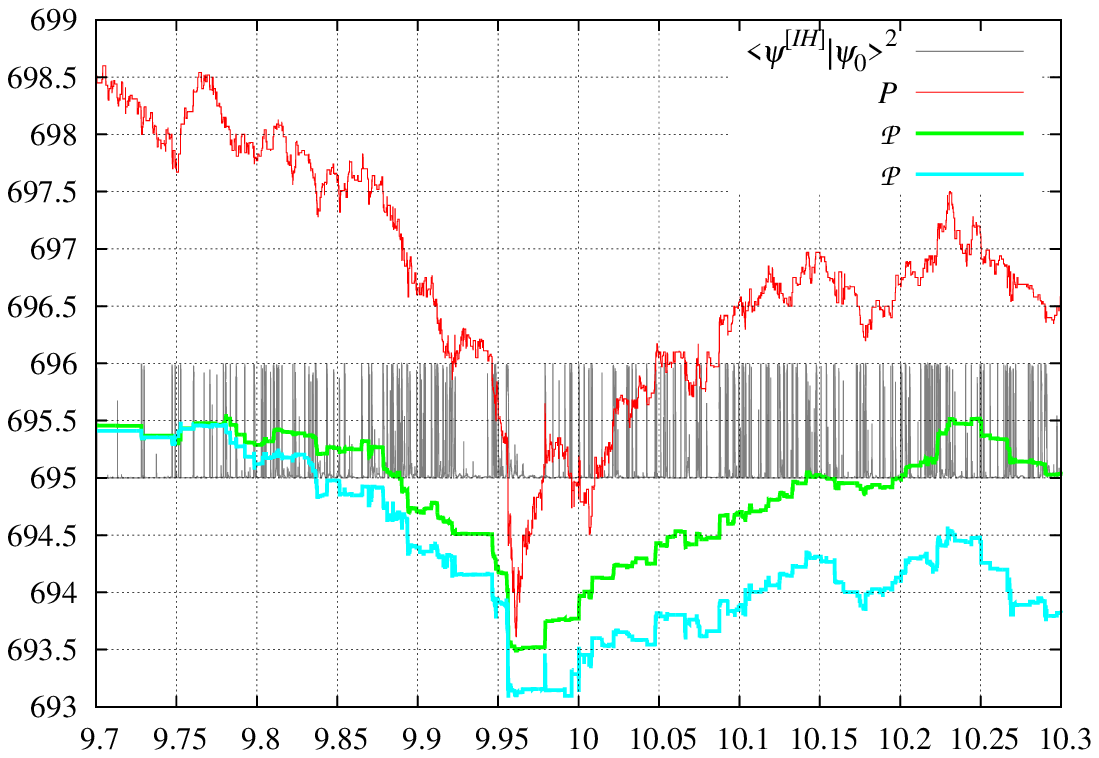}
  \caption{\label{PPCompare}
    The comparison of scalp--price ${\cal P}$ obtained from ${\cal F}_l$:
    from (\ref{DirFactProxy}) (green: \texttt{{.dp\_to\_use}=F\_dpdt0\_SCALP})
    from (\ref{DirFactSimple}) (blue: \texttt{{.dp\_to\_use}=F\_SAMPLE\_DP\_SCALP}).
    The $\Braket{\psi_I^{[IH]}|\psi_0}^2$ is used as
    a scalp--function ${\cal S}(t)$ (\ref{Wdef}).
    The scalp--prices are shifted
    to fit the chart;  they are defined (\ref{PriceFromdF}) within a constant.
    If one use \texttt{{.dp\_to\_use}=F\_SAMPLE\_DP\_NOSCALP} (\ref{Fleqdp})
    the result will be exactly 
    the price $P$,
    shifted by some initial level.
  }
\end{figure}

Output file is a tab--separated file
with the columns (35 columns total), corresponding to the results of this paper.
Field names are printed in the first line of the output file.
The data can be processed by any common plotting software (such as gnuplot or matlab).
Below is the description of the most noticeable fields:
\begin{itemize}
\item \verb+T+ : Time in nanoseconds since midnight (copied from input).
\item \verb+shares+ : Shares traded (copied from input).
\item \verb+P_last+ : Last execution price (copied from input).
\item \texttt{\seqsplit{\{pi\_average,pt\_average\}}} :
  Regular exponential moving average of price
  with the given \texttt{\seqsplit{--tau=128}}, using volume/time as the weight.
\item \texttt{\seqsplit{I.\{s0,sL,wL\_squared,sH,wH\_squared,Gamma0\}}} :
  Correspond to (\ref{EVXData:I:fields}) fields
  of $\Ket{I|\psi}=\lambda\Ket{\psi}$ eigenvalue problem (\ref{GEVI}), the solution 
  with the given \texttt{\seqsplit{--n=12}}; the \texttt{\seqsplit{I.wL}} and \texttt{\seqsplit{I.wH}} are squared in the output,
  $\mathtt{Gamma0}=\left(2I_0-\lambda_{I}^{[IL]}-\lambda_{I}^{[IH]}\right)\Big/\left(\lambda_{I}^{[IL]}-\lambda_{I}^{[IH]}\right)$
  is the $\widetilde{\Gamma^0}$ skewness of $I$, Eq. (\ref*{Mu-skewnesslikeS0})
  of Ref. \cite{ArxivMalyshkinMuse}. The \texttt{\seqsplit{I.wH\_squared}}
  is the scalp--function ${\cal S}(t)$ (\ref{WprojIH}).
\item \texttt{\seqsplit{\{p\_IH,pt\_IH,pV\_IH,pT\_IH\}}}
  Correspond to (\ref{pvtdev:withoutSuccess}) prices, calculated in the
  state $\Ket{\psi_I^{[IH]}}$  (\ref{rhoIH}), the (\ref{ScalpedMaxIProjection:fields}) fields.
\item \texttt{\seqsplit{getFlFromRegularMoments()}} The ${\cal F}_l$ when it is
  calculated from regular moments, \texttt{\seqsplit{NaN}} otherwise,
  the field (\ref{getFlFromRegularMoments:field}).
\item \texttt{\seqsplit{getSumFdt()}}
  The scalp--price ${\cal P}$ (\ref{PriceFromdF}),
  corresponding to given \texttt{\seqsplit{dp\_to\_use}},
  the field (\ref{scalpPrice:field}).
  See the Fig. \ref{PPCompare} to compare the results
  for \texttt{\seqsplit{.dp\_to\_use=F\_dpdt0\_SCALP}} (\ref{DirFactProxy}) and \texttt{\seqsplit{.dp\_to\_use=F\_SAMPLE\_DP\_SCALP}} (\ref{DirFactSimple}).
\item \texttt{\seqsplit{dIH,dp\_IH}} The $\lambda_I^{[IH]}$ and $p^{[IH]}$
  change \textsl{per tick}, the fields (\ref{scalpPrice:fielddIG})
  and (\ref{scalpPrice:fielddpIG}). This is the starting point of non--local
  price change (\ref{dptodFpIH}) study, Fig. \ref{FlPIHExampleFig}.
\item \texttt{\seqsplit{\{DIR,DIRa\}}} and etc.
  Correspond to (\ref{ScalpedMaxIProjection:fields}) fields of an object of
  \hyperref[ScalpedMaxIProjection]{\texttt{\seqsplit{ScalpedMaxIProjection.java}}} type.
\end{itemize}

\subsection{\label{RunProgramUsage}Installation and usage example}
\begin{itemize}
\item Install java 1.8 or later.
\item
  Download
  from \cite{polynomialcode}
   the archive
\href{http://www.ioffe.ru/LNEPS/malyshkin/AMuseOfCashFlowAndLiquidityDeficit.zip}{\texttt{\seqsplit{AMuseOfCashFlowAndLiquidityDeficit.zip}}}
  with the source code.
\item Decompress and recompile the program:
\begin{verbatim}
unzip AMuseOfCashFlowAndLiquidityDeficit.zip
javac -g com/polytechnik/*/*java
\end{verbatim}

\item Run the test with the bundled file
  \texttt{\seqsplit{dataexamples/aapl\_old.csv.gz}}
  data of Ref. \cite{2016arXiv160204423G}. The file contains
 only execution events,
  the (time, execution price, shares traded)
  market observations triples are in the 1:2:3 columns, column index is base 0;
  28492 lines, 9 columns total.
\begin{verbatim}
java com/polytechnik/algorithms/CallAMuseOfCashFlowAndLiquidityDeficitWithScalp \
      --musein_file=dataexamples/aapl_old.csv.gz \
      --musein_cols=9:1:2:3 \
      --museout_file=museout.dat \
      --n=12 \
      --tau=128 \
      --measure=ScalpedMaxIProjectionLegendreShifted
\end{verbatim}
The code is run under 16 seconds,
the output fields of the \texttt{\seqsplit{museout.dat}} are described in
the Appendix \ref{CallAMusestucture}. The
\texttt{\seqsplit{I.wH\_squared}},
\texttt{\seqsplit{getSumFdt()}}, and
\texttt{\seqsplit{p\_IH}} are the scalp--function (\ref{WprojIH}),
scalp--price (\ref{PriceFromdF}) (has an arbitrary offset),
and $p^{[IH]}$ from (\ref{pIH}).
The default \texttt{\seqsplit{.dp\_to\_use=F\_PROBABILITYCORRELATION\_SCALP}}
corresponds to (\ref{FlProbabilityCorrelation2DirSignedScalp}).
\item
   Download
NASDAQ ITCH data file
\texttt{\seqsplit{S092012-v41.txt.gz}}
from \cite{polynomialcode},
extract triples (time, execution price, shares traded)
from NASDAQ ITCH data file:
\begin{verbatim}
java com/polytechnik/itch/DumpData2Trader \
      S092012-v41.txt.gz AAPL >aapl.csv
\end{verbatim}
Execution data and limit order book edges are now saved to tab--separated file \verb+aapl.csv+
of 15 columns. The (time, execution price, shares traded)
market observations triples are in the 1:4:5 columns, column index is base 0;
  634205 lines, 15 columns total.
\item Run the java command of the Appendix \ref{CallAMusestucture}
  to obtain the \texttt{\seqsplit{museout.dat}}
  file of 634206 lines with: scalp--function (\ref{WprojIH}),
scalp--price (\ref{PriceFromdF})
and $p^{[IH]}$ from (\ref{pIH})
and
\hyperref[ScalpedMaxIProjection:fields]{\texttt{\seqsplit{com/polytechnik/trading/ScalpedMaxIProjection.java}}}
fields is created. The code is run under 5 minutes,
much longer than that of previous run.
The \texttt{\seqsplit{--musein\_file=aapl.csv}} input file
now contains much more events
than the file \texttt{\seqsplit{--musein\_file=dataexamples/aapl\_old.csv.gz}}.
  
\end{itemize}

\section{\label{VdivT}The state of maximal aggregated execution flow $V/t$}
In our previous work\cite{2015arXiv151005510G,ArxivMalyshkinMuse}
the extremal state of $I=dV/dt$ operator have been considered.
This answer has two critically important features:
\begin{itemize}
\item Uses execution flow $I$, as it is the driving force of the market.
\item Has automatic time--scale selection (eigenvalue problem),
  huge advantage compared to any fixed time scale approach\cite{2015arXiv151005510G}.
\end{itemize}
While this result is very promising,
it has an issue of zero first variation of $I$.
Consider the same approach, but with the operator
$V/t$. Here $V$ and $t$ are measured \textsl{since $t_{now}$},
they are  volume/time between $t$ and $t_{now}$.
The $V/t$ is \textbf{aggregated} execution flow, the $dV/dt$
is \textbf{local} execution flow.
Put $f=V/t$ into (\ref{GEV}) and obtain
generalized eigenvalue problem to find
the state $\Ket{\psi_{V/t}^{[\max]}}$
of maximal $\lambda_{V/t}^{[\max]}$:
\begin{eqnarray}
  \Ket{V\Big|\psi_{V/t}^{[i]}}&=&\lambda_{V/t}^{[i]}\Ket{t\Big|\psi_{V/t}^{[i]}} \label{GEVdefVT} \\
\sum\limits_{k=0}^{n-1} \Braket{Q_j|V|Q_k} \alpha^{[i]}_k &=&
  \lambda_{V/t}^{[i]} \sum\limits_{k=0}^{n-1} \Braket{ Q_j|t|Q_k} \alpha^{[i]}_k
  \label{GEV_VT} \\
  \psi_{V/t}^{[i]}(x) &=& \sum\limits_{k=0}^{n-1} \alpha^{[i]}_k Q_k(x) \label{psi_GEV_VT}
\end{eqnarray}
The calculation of $\Braket{Q_j|V|Q_k}$ and $\Braket{ Q_j|t|Q_k}$
matrix elements is described in the Appendix \ref{VMatrixElements}.
In (\ref{GEV_VT}) the $V$ and $t$ have the sign changed
to have positively defined right--hand--side matrix $\Braket{ Q_j|t|Q_k}$,
$V=V_0$ (\ref{cmdefV}), $t=T_0$ (\ref{cmdefT}).
The multiplication by $V$ and $t$ create, for $t\le t_{now}$,
two
\href{https://www.encyclopediaofmath.org/index.php/Radau_quadrature_formula}{Radau--like measures}:
$\left(V(t_{now})-V(t)\right)\omega(t)dt$ and
$\left(t_{now}-t\right)\omega(t)dt$.
The problem (\ref{GEV_VT}) finds the state $\psi_{V/t}^{[\max]}(x)$,
corresponding to the maximal Radon--Nikodym derivative
relatively
two these measures, the maximal aggregated execution flow $V/t$.
Previously \cite{2015arXiv151005510G}
we have been considering the state $\psi_{I}^{[IH]}(x)$,
corresponding to the maximal Radon--Nikodym derivative
relatively the measures $\omega(t)dV$ and $\omega(t)dt$, the maximal local execution flow $dV/dt$.
The eigenvectors $\Ket{\psi_{V/t}^{[i]}}$ of $\|V/t\|$ operator
have the following remarkable features:
\begin{align}
  \intertext{Normalized to Radau--like measure $\left(t_{now}-t\right)\omega(t)dt$:}
  1&=\Braket{\psi_{V/t}^{[i]}|t|\psi_{V/t}^{[i]}} \label{VTnorm}\\
  \intertext{In the $\Ket{\psi_{V/t}^{[i]}}$
states aggregated $V/t$ and local $dV/dt$ execution flows are equal:}
  \lambda_{V/t}^{[i]}&=
  \frac{\Braket{\psi_{V/t}^{[i]}|V|\psi_{V/t}^{[i]}}}
       {\Braket{\psi_{V/t}^{[i]}|t|\psi_{V/t}^{[i]}}}=
       \frac{\Braket{\psi_{V/t}^{[i]}|I|\psi_{V/t}^{[i]}}}
            {\Braket{\psi_{V/t}^{[i]}|\psi_{V/t}^{[i]}}} \label{lambdaEq} \\
            \intertext{For infinitesimal time--shift $\delta \psi =D(\psi_{V/t}^{[i]})$
              the second variation (\ref{rqD2}) of $V/t$
              is equal to the first variation (\ref{rqD1}) of $dV/dt$:}
      &      \Braket{\delta \psi | V | \delta \psi}-
            \lambda_{V/t}^{[i]}\Braket{\delta \psi | t | \delta \psi}
            =
            \Braket{\delta \psi | I | \psi_{V/t}^{[i]}}
            -\lambda_{V/t}^{[i]}\Braket{\delta \psi | \psi_{V/t}^{[i]}}
            \label{var2VTeqvar1I}
\end{align}
\begin{lem}
In the state of
maximal aggregated execution flow the $dI/dt$ is positive.
\begin{proof}
  In the state of
  maximal $V/t$ the second variation (\ref{rqD2}) is negative.
  Because the first $I$ variation (\ref{rqD1}) with $\delta \psi =D(\psi_{V/t}^{[i]})$
  corresponds to $-dI/dt$, this provides positive $dI/dt$.  
\end{proof}
\end{lem}
This lemma makes the state $\Ket{\psi_{V/t}^{[\max]}}$
of maximal aggregated execution flow (the eigenvector of (\ref{GEV_VT}),
corresponding to the maximal $\lambda_{V/t}^{[\max]}$),
a very promising one for the market dynamics to consider.
The ``aggregated''  attributes (\ref{inputmatricesAggregated})
have been originally introduced
in the Section (\ref*{Mu-PsiAfterFutureI}) ``Measure: The Period After Maximal Future $I$'' of Ref. \cite{ArxivMalyshkinMuse},
but their application to skewness study was not a very successful back then.

\section{\label{VMatrixElements}The calculation of the $\Braket{VQ_m}$
  moments from the $\Braket{IQ_m}$ moments}
The $\Braket{VQ_m}$ and $\Braket{tQ_m}$, $m=[0\dots 2n-2]$,
moments, required to construct $\|V\|$ and $\|t\|$
operators in (\ref{GEVdefVT}),
can be calculated directly from the sample. However,
in practical application it is more convenient
to calculate the $\Braket{IQ_m}$ moments first,
then to obtain the $\Braket{VQ_m}$ moments using an integration by parts.
For Shifted Legendre and Laguerre bases the integration by parts gives:
\begin{eqnarray}
  \int_{-\infty}^{t_{now}}VQ_m(x(t))\omega(t)dt&=&
  V(t_{now})Q_m(x(t_{now}))-\int_{-\infty}^{t_{now}}J(Q_m(x(t)))\omega(t)Idt
  \label{VbyParts}
\end{eqnarray}
where  $J(\cdot)$ is a
polynomial to polynomial transforming function (\ref{wfPartsdef}).
The $\Braket{VQ_m}$ then can be expressed as $\Braket{IQ_s}$, $s=[0\dots m]$,
linear combination. This is possible \textsl{only} for the bases
in question, in general case an integration by parts
$\int_{-\infty}^{t} Q_m(x(t^{\prime}))\omega(t^{\prime})dt^{\prime}$ cannot be reduced
to a $J(Q_m(x(t)))\omega(t)$ form, and
the $\Braket{VQ_m}$ moments
cannot be expressed via a linear combination of the $\Braket{IQ_s}$ moments.

The boundary condition is straightforward,
consider $V(t)-V(t_{now})$, that is zero at $t=t_{now}$.
Use current volume $V(t_{now})$ as the starting value,
then out--of--integral term in (\ref{VbyParts}) vanish,
and past/future volume correspond to negative/positive volume values\footnote{
  It is sometimes convenient to change the sign of time and volume $V(t)-V(t_{now})$
  as in (\ref{inputmatricesAggregated}),
  then past time and volume correspond to positive
  values and the right hand side matrix in (\ref{GEVdefVT})
  is positively defined.  
  }.
See the method \texttt{\seqsplit{setFMoments}} of
\texttt{\seqsplit{com/polytechnik/trading/\{QVMDataLDirectAccess,QVMDataPDirectAccess,QVMDataDirectAccess\}.java}},
that calculates the $\Braket{VQ_m}$ moments
as a linear combination of the $\Braket{IQ_s}$, $s=[0\dots m]$ moments.

\section{\label{dIdtMatrixElements}The calculation of the $\|dI/dt\|$
  operator matrix elements from the $\|I\|$ operator.}
When we study an operators of execution rate change $\|dI/dt\|$,
it's matrix elements  cannot be calculated
directly from sample.
In general case the $\Braket{\frac{dI}{dt}Q_m}$ moments
can be calculated from $\Braket{IQ_m}$ moments using
integration by parts (\ref{VbyParts}) of the  Appendix \ref{VMatrixElements},
see the method \texttt{\seqsplit{setDFMoments}} of
\texttt{\seqsplit{com/polytechnik/trading/\{QVMDataLDirectAccess,QVMDataPDirectAccess,QVMDataDirectAccess\}.java}},
that, for zero boundary condition,
obtains $\Braket{\frac{dI}{dt}Q_m}$
as a linear combination of  $\Braket{IQ_s}$, $s=[0\dots m]$.
However, for $\|dI/dt\|$, the boundary condition
may take a variety of forms, and direct operator approach
is often more convenient.
 Consider e.g. generalized eigenvalue problem (\ref{GEV}) for $\|dI/dt\|$ operator:
\begin{eqnarray}
  \Ket{\frac{dI}{dt}\Big|\psi_{\frac{dI}{dt}}^{[i]}}&=&
  \lambda_{\frac{dI}{dt}}^{[i]}\Ket{\psi_{\frac{dI}{dt}}^{[i]}}
  \label{GEVdI}
\end{eqnarray}
where the $\Braket{Q_j\left|\frac{dI}{dt}\right|Q_k}$ matrix cannot be calculated directly
from sample.
For a $Q_k(x)$ basis with infinitesimal time--shift operator $D(Q_k(x))$, the matrix
can be obtained from the $\Braket{Q_j\left|I\right|Q_k}$ matrix
using integration by parts\footnote{
  See \href{http://www.ioffe.ru/LNEPS/malyshkin/AMuseOfCashFlowAndLiquidityDeficit.zip}{java classes} for Shifted Legendre and Laguerre $Q_k(x)$ bases
  implementation
  of infinitesimal time--shift operator $D(Q_k(x))$:
  the method \texttt{\seqsplit{getEDPsi}} of
  \texttt{\seqsplit{com/polytechnik/trading/\{WIntegratorLegendreShifted,WIntegratorLaguerre,WIntegratorMonomials\}.java}}. Also see the \texttt{\seqsplit{com/polytechnik/trading/QQdidtMatrix.java}} class, implementing the calculation of (\ref{dIMatr}) matrix for the
  (\ref{BOUNDARY:IeqIH}),
  (\ref{BOUNDARY:DIeq0}),
  (\ref{BOUNDARY:IeqI0}),
  and (\ref{BOUNDARY:Ieq0}),
  boundary conditions.
   This class uses \texttt{\seqsplit{com/polytechnik/utils/VolMatrix.java}}
   to calculate $\Braket{D(Q_j)|I|Q_k}+\Braket{Q_j|I|D(Q_k)}$,
   then adds boundary condition term $I^{f}Q_j(x_0)Q_k(x_0)$.
}, Eq. (\ref*{Mu-dIMatr}) of Ref. \cite{ArxivMalyshkinMuse}:
\begin{eqnarray}
  \Braket{Q_j\left|\frac{dI}{dt}\right|Q_k} &=& I^{f}Q_j(x_0)Q_k(x_0)- \Braket{D(Q_j)|I|Q_k}-\Braket{Q_j|I|D(Q_k)}
  \label{dIMatr}
\end{eqnarray}
This problem is an inverse one to considered in Appendix \ref{VMatrixElements}, and requires
a non--trivial boundary condition $I^{f}$.
There are several options for $I^{f}$, that  can be reasonably considered:
\begin{itemize}
\item The zero of $\|dI/dt\|$ in the $\Ket{\psi_{I}^{[IH]}}$ state,
$\Braket{\psi_{I}^{[IH]}|\frac{dI}{dt}|\psi_{I}^{[IH]}}=0$:
\begin{align}
  I^{f}&= \lambda_I^{[IH]}
  \label{BOUNDARY:IeqIH}
\intertext{
\item The zero of $\|dI/dt\|$ in the $\Ket{\psi_0}$ state,
  $\Braket{\psi_0|\frac{dI}{dt}|\psi_0}=0$:}
  I^{f}&= 2\frac{\Braket{\psi_0|I|D(\psi_0)}}{\psi^2_0(x_0)}
  \label{BOUNDARY:DIeq0}
\intertext{
\item The $I_0$ value:}
  I^{f}&= \Braket{\psi_0|I|\psi_0}
  \label{BOUNDARY:IeqI0}
\intertext{
\item Zero value:
  }
  I^{f}&= 0
  \label{BOUNDARY:Ieq0}
\end{align}
\end{itemize}
Regardless the $I^{f}$ selection,
the $\|I\|$ and $\|dI/dt\|$ operators have no common
eigenvectors unless the $\Ket{\psi_0}$ is the $\|I\|$ eigenvector,
this degeneracy case was considered in Ref. \cite{ArxivMalyshkinMuse}.
The most critical degeneracy arise in the situation,
when the state ``now'' and the state of ``maximal past $I$'' are the same:
\begin{eqnarray}
  \Ket{\psi_0}&=&\Ket{\psi_{I}^{[IH]}}
  \label{degeneracyPsi0PsiIH}
\end{eqnarray}
An example of such a degeneracy can be
the situation of huge volume traded ``now'' (at $x=x_0$).

\section{\label{PnLRatio}Directional Information:
  $I\xrightarrow[{\psi}]{\quad }\max$
  Subject To the  Constraint $\Braket{\psi|C|\psi}=0$.
}
 Consider market dynamics split in two operators:
$\|I\|$ (execution flow dynamics)
and $\|C\|$ (price dynamics).
The constrained $I\to\max$ problem is:
\begin{subequations}
  \label{ImaxConstarined}
\begin{align}
I&=\frac{\Braket{\psi|I|\psi}}{\Braket{\psi|\psi}} \xrightarrow[{\psi}]{\quad }\max
  \label{ImaxconstainedMax} \\
\text{subject to:}\quad
0&=\Braket{\psi|C|\psi} \label{ImaxPPlconstaint}
\end{align}
\end{subequations}
The constraint (\ref{ImaxPPlconstaint})  is
a requirement on price in the $\Ket{\psi}$ state.
There are a number of choices for the constraint operator $\|C\|$ selection:
\begin{subequations}
  \label{ConstraintsSelection}
\begin{align}
  \|C\|&=\|\left(p-P^{last}\right)I\| \label{constraintPriceEqPlastDef}
  & &\text{Price (\ref{pvDefine}) in the $\Ket{\psi}$ state is equal to $P^{last}$}  \\
   \|C\|&=\|V_1-P^{last}V_0\| \label{constraintPriceMovingAverageEqPlastDef}
   & &\text{Moving average price (\ref{pVDefine}) is equal to $P^{last}$}  \\
   \|C\|&=\left\|\frac{d}{dt}\left[\left(p-P^{last}\right)I\right]\right\| \label{constraintPricedIEqPlastDef}
   & &\text{Price--execution flow changes match} \\
   \|C\|&=\left\|\frac{dp}{dt}\right\| \label{constraintPriceExtremum}
   & & \text{Price extremum}  \\
   \|C\|&=\left\|\frac{d^2p}{dt^2}\right\| \label{constraintDPExtremum}
   & & \text{$dp/dt$ extremum} 
\end{align}
\end{subequations}
The maximization problem (\ref{ImaxconstainedMax}) with the quadratic constraint
(\ref{ImaxPPlconstaint})
can no longer be reduced to a regular
eigenvalue problem such as (\ref{GEVI}).
The solution exists only if $\|C\|$
operator has both: positive and negative eigenvalues.
Ideologically the (\ref{ImaxPPlconstaint}) constraint
facilitate
taking into account a typical market practitioner activity:
look how the market used to behave in the past
at prices near some level.
Our previous paper \cite{ArxivMalyshkinMuse}
has been mostly devoted to skewness and probability correlation
study
in the unconstrained $I\to\max$ state $\Ket{\psi_I^{[IH]}}$.
The (\ref{ImaxPPlconstaint}) constraint
allows us,
within the framework of a single formalism
of constrained optimization,
take into account
 the
driving force of the market  $I\to\max$ (\ref{ImaxconstainedMax})
and the reaction of the market participants on it (\ref{ImaxPPlconstaint}).
For mathematical properties and numerical solution
of (\ref{ImaxConstarined}) problem see
Appendices \ref{IstatesConditional}
and (\ref{IstatesConditionalLocalized}
below.
Here we assume that the solution does exist,
we denote it as $\Ket{\psi_I^{[{\cal M}]}}$,
and name: the state of price--matching maximal execution flow.
The found state $\Ket{\psi_I^{[{\cal M}]}}$ (it is just a pure state averaging weight
$\left(\psi_I^{[{\cal M}]}(x(t))\right)^2\omega(t)dt$, not even a density matrix (\ref{densmatrdef}))
is the state to obtain market directional information.

\subsection{\label{IstatesConditional}The \texttt{IstatesConditional.java} implementation}
The optimization problem (\ref{ImaxconstainedMax}) with quadratic
constraint (\ref{ImaxPPlconstaint}) can  be solved using Lagrange multipliers
technique:
\begin{subequations}
  \label{lagrangemultAllEqs}
\begin{align}
  \max\limits_{\psi}&\;\Braket{\psi|I|\psi} -\lambda(\Braket{\psi|\psi}-1)+
  \mu\Braket{\psi|C|\psi}
  \label{lagrangemultMax} \\
  1 &=\Braket{\psi|\psi} \label{lagrangemultpsinorm} \\
  0 &=\Braket{\psi|C|\psi}
  \label{lagrangemultquadrconstr} \\
  \Ket{0} &=\Ket{I|\psi} -\lambda\Ket{\psi}+\mu\Ket{C|\psi}
  \label{lagrangemultvar}
\end{align}
\end{subequations}
Were the constraint (\ref{lagrangemultquadrconstr})
to be of a linear type, instead of a quadratic one,
the constrained optimization problem (\ref{lagrangemultMax})
can be reduced to a regular eigenvalue problem
in a transformed basis\cite{golub1973some}.
However, for the quadratic constraint (\ref{lagrangemultquadrconstr}),
such a one--step transform is not possible,
and self--concordant procedure of iterational type is the simplest option:
\begin{itemize}
\item For an initial $\Ket{\psi}$ find the coefficient $\alpha$, such that:
  \begin{subequations}
    \label{optc}
    \begin{align}
      \Ket{b}&=\Ket{C|\psi} \label{badj}\\
      0 &=\Braket{\psi+\alpha b|C|\psi+\alpha b} \label{badjvar}
  \end{align}
  \end{subequations}
The (\ref{badjvar}) is a quadratic equation with respect to $\alpha$,
if no real solution exist --- iterational process failed. If a success ---
obtain the solution,
satisfying the (\ref{lagrangemultquadrconstr})
constraint:
\begin{align}
  \Ket{\widetilde{\psi}}&=\Ket{\psi} +\alpha \Ket{b}
  \label{psiadjustedtobadj}
\end{align}
From the two
$\alpha$ solutions select the one with the maximal
${\Braket{\widetilde{\psi}|I|\widetilde{\psi}}}\Big/{\Braket{\widetilde{\psi}|\widetilde{\psi}}}$.
There are exist several good alternatives to (\ref{optc}),
see 
\texttt{\seqsplit{com/polytechnik/utils/IstatesConditional.java}}
implementation for details.
\item  Put $\Ket{\widetilde{\psi}}$ to (\ref{lagrangemultvar}),
then left--multiply it by the vector
$\Bra{\widetilde{\psi}\Big|C}$, obtain the Lagrange
multiplier iteration $\mu$:
\begin{align}
  \mu &=-\,\frac{\Braket{\widetilde{\psi}|C|I|\widetilde{\psi}}}
      {\Braket{\widetilde{\psi}|C|C|\widetilde{\psi}}}
      \label{lagrangemultiplierMuDef}
\end{align}
\item
Construct an operator $\|{\cal I}\|$ and find all it's eigenvectors:
\begin{align}
\|{\cal I}\| &= \|I\| + \mu \|C\| \label{Ilagmult} \\
   \Ket{{\cal I}\Big|\psi^{[i]}}&=\lambda^{[i]}\Ket{\psi^{[i]}}
   \label{GEVIlagmult}
\end{align}
\item Among all the $\Ket{\psi^{[i]}}$ found
  select
  the $\Ket{\psi}$,
  providing the maximal $\Braket{\psi|I|\psi}$.
\item
 Repeat the process of above for this new $\Ket{\psi}$.
  If a solution exists,
  iterational procedure converges quickly (typically 5--7 iterations),
  unless $\|I\|$ and $\|C\|$ operators
  have several eigenvectors in common\footnote{
    Assume $\|I\|$ and $\|C\|$ operators
    have the identical eigenvectors. Then
    the (\ref{GEVIlagmult}) always produce the same
    eigenvectors, and the minimization problem (\ref{ImaxConstarined})
    is reduced to a \href{https://en.wikipedia.org/wiki/Linear_programming}{linear programming} problem relatively
    the projections squares.
    }.
  The result of this iterational process is
  the state of price--matching maximal execution flow
  $\Ket{\psi_I^{[{\cal M}]}}$,
  the (\ref{ImaxConstarined}) solution.
\end{itemize}

The class
\texttt{\seqsplit{com/polytechnik/utils/IstatesConditional.java}}
implements this procedure.
It 
takes  three matrices $\Braket{Q_j|Q_k}$, $\Braket{Q_j|I|Q_k}$,
$\Braket{Q_j|C|Q_k}$,
and basis functions operations class (extending the
\texttt{\seqsplit{com/polytechnik/utils/OrthogonalPolynomialsABasis.java}}),
as constructor's arguments.
Then it solves generalized eigenvalue problem (\ref{GEVI})
using the \hyperref[EVXData]{\texttt{\seqsplit{EVXData.java}}} class
to obtain an initial $\Ket{\psi}$ and to reproduce
the \cite{2015arXiv151005510G} results.
Then ten iterations of above
are performed
to obtain the solution of (\ref{ImaxConstarined}): $\Ket{\psi_I^{[{\cal M}]}}$ and $\mu$.
The fields are:
\begin{subequations}
  \label{ImaxConstarined:fields}
  \begin{align}
    &\mathtt{.I}& &  \text{An object of \hyperref[EVXData]{\texttt{\seqsplit{EVXData.java}}} type, (\ref{GEVI}) solution} \label{liquiditydeficitI}\\
    &\mathtt{.flag\_solution\_exists} &&
    \text{Whether the (\ref{ImaxConstarined}) solution exists for the input data} \\
    &\mathtt{.psi\_M}& &\text{$\Ket{\psi_I^{[{\cal M}]}}$ the (\ref{ImaxConstarined}) solution; equals to 0 on failure} \label{psi:M}\\
    &\mathtt{.LagrangeMultiplier\_M}& &\text{Lagrange multiplier $\mu$, Eq. (\ref{lagrangemultiplierMuDef})} \label{labrangemultiiplier} \\
    &\mathtt{.i\_M}& &\text{$\Braket{\psi_I^{[{\cal M}]}|I|\psi_I^{[{\cal M}]}}$ execution flow in the $\Ket{\psi_I^{[{\cal M}]}}$ state} \\
    &\mathtt{.wr0\_M}& & \text{$\Braket{\psi_I^{[{\cal M}]}|\psi_0}^2$ a kind of ``distance to now''}
  \end{align}
\end{subequations}

\subsection{\label{IstatesConditionalLocalized}The \texttt{IstatesConditionalLocalized.java} implementation}
When the global maximum of constrained $I\to\max$ problem
is not required, and localized answer with $\Ket{\psi}$ in (\ref{psiRN})
form is considered as good enough,
optimization problem (\ref{ImaxconstainedMax}) with quadratic
constraint (\ref{ImaxPPlconstaint}) can  be easily solved.
Substitute  (\ref{psiRN}) to (\ref{lagrangemultquadrconstr}) and obtain:
\begin{subequations}
  \label{ImaxLocalizedAllEqs}
\begin{align}
    I&=\frac{\Braket{\psi_y|I|\psi_y}}{\Braket{\psi_y|\psi_y}}
     \xrightarrow[{y}]{\quad }\max
  \label{ImaxLocalized}\\
  0&=
  \sum\limits_{j,k,s,t=0}^{n-1} Q_j(y)G^{-1}_{jk}\Braket{Q_k|C|Q_s}G^{-1}_{st}Q_t(y)
  \label{polynomial0equation}
\end{align}
\end{subequations}
The (\ref{polynomial0equation}) constraint is a polynomial of $2n-2$ degree, it has
exactly $2n-2$ root, possibly complex. The classes extending
the \texttt{\seqsplit{com/polytechnik/trading/OrthogonalPolynomialsABasis.java}}
(see Appendix \ref*{Mu-codestr} of Ref. \cite{ArxivMalyshkinMuse})
provide an implementation for solving $P(y)=0$ equation
with a $P(y)$ in a given $Q_k(y)$
basis $P(y)=\sum_{m=0}^{2m-2}Q_m(y)$, the (\ref{polynomial0equation}) is
a polynomial of this form.
Among $2n-2$ roots found select only the real roots, then among them
select the state $\Ket{\psi_I^{[{\cal M}]}}$, that provides
the maximal $\Braket{\psi_I^{[{\cal M}]}|I|\psi_I^{[{\cal M}]}}$.
The situation is similar to the one of Appendix \ref{ConstrainsOperator},
below, with the difference that $\Ket{\psi_I^{[{\cal M}]}}$
is now selected among (\ref{psiRN}) states with $y$
from (\ref{polynomial0equation}) real roots ($2n-2$ maximal number),
not among $n$ eigenvalues of some operator $\|{\cal C}\|$.

The class
\texttt{\seqsplit{com/polytechnik/utils/IstatesConditionalLocalized.java}}
implements this procedure.
It 
takes  three matrices $\Braket{Q_j|Q_k}$, $\Braket{Q_j|I|Q_k}$,
$\Braket{Q_j|C|Q_k}$,
and basis functions operations class (extending the
\texttt{\seqsplit{com/polytechnik/utils/OrthogonalPolynomialsABasis.java}}),
as constructor's arguments.
Then it solves $P(y)=0$ polynomial roots problem (\ref{polynomial0equation})
using \texttt{\seqsplit{com/polytechnik/trading/OrthogonalPolynomialsABasis:getPolynomialRootsFinderInBasis().findRoots($\cdot$)}} method
to obtain a set of $y_m$ that are the roots of (\ref{polynomial0equation}).
Then corresponding $\Ket{\psi_{y_m}}$ (\ref{psiRN}) are constructed,
and the one with the maximal $\Braket{\psi_{y_m}|I|\psi_{y_m}}$
is selected: this is the ``localized'' $\Ket{\psi_I^{[{\cal M}]}}$ solution.
The fields are:
\begin{subequations}
  \label{ImaxConstarinedLocalizd:fields}
  \begin{align}
    &\mathtt{.psi\_M}& &\text{$\Ket{\psi_I^{[{\cal M}]}}$ the (\ref{ImaxConstarined}) localized solution of (\ref{psiRN}) form; equals to 0 on failure} \label{psi:M:Localized}\\
    &\mathtt{.y\_M}& &  \text{The ``localization'' point in (\ref{psiRN}) of maximal $I$ (\ref{ImaxLocalized}),
(\ref{polynomial0equation}) root} \label{y:M:Localized}\\
    &\mathtt{.i\_M}& &\text{The execution flow $\Braket{\psi_I^{[{\cal M}]}|I|\psi_I^{[{\cal M}]}}$} \label{i:M:Localized}\\
    &\mathtt{.n\_roots} & & \text{The number of real roots of (\ref{polynomial0equation})}
    \label{nroots:M:Localized}
  \end{align}
\end{subequations}

\section{\label{ConstrainsOperator}Directional Information:
  $I\xrightarrow[{\psi}]{\quad }\max$
  in the States of Constraint Operator $\|{\cal C}\|$.
}
The constrained optimization of the Appendix \ref{IstatesConditional}
above,
while been very nice mathematically,
does not provide a clear cut answer.
There are two reasons: the difficulty to select
an operator $\|C\|$ (\ref{ConstraintsSelection})
and the difficulty
with (\ref{GEVIlagmult}) Lagrange multiplier convergence,
as $\|I\|$ and $\|C\|$ operators often have common eigenvectors.
Consider a different, much more simplistic, constrained optimization approach:
\begin{subequations}
  \label{ImaxConstarinedSelection}
\begin{align}
I&=\frac{\Braket{\psi|I|\psi}}{\Braket{\psi|\psi}} \xrightarrow[{\psi}]{\quad }\max
  \label{ImaxconstainedMaxSelection} \\
\Ket{\psi}&: \text{is subject to being an eigenvector of $\Ket{{\cal C}|\psi}=\lambda_{{\cal C}}\Ket{\psi}$ } \label{ImaxPPlconstaintSelection}
\end{align}
\end{subequations}
Here we also split
the market dynamics in two operators:
$\|I\|$ (execution flow dynamics)
and $\|{\cal C}\|$ (price dynamics).
But now we consider the $\|I\|$ only in the eigenstates
of the operator $\|{\cal C}\|$.
The operator $\|{\cal C}\|$ is selected in a way
that it's derivative gives the constraint operator $\|C\|$,
thus the $\Ket{\psi}$ state of extremal $\|{\cal C}\|$ give zero of
constraint operator $\|C\|$.
Mathematically the problem (\ref{ImaxConstarinedSelection}) is simple:
find all $n$ eigenvectors (\ref{ImaxPPlconstaintSelection}) of
$\|{\cal C}\|$ first,
then select the one, providing the maximal $\|I\|$ (\ref{ImaxconstainedMaxSelection}).
The state of price--matching maximal execution flow
$\Ket{\psi_I^{[{\cal M}]}}$ is now plain (\ref{ImaxPPlconstaintSelection})
eigenvector, providing the maximal (\ref{ImaxconstainedMaxSelection}).
There are a number of choices for the operator $\|{\cal C}\|$,
selecting the states $\Ket{\psi}$:
\begin{subequations}
  \label{ConstraintsSelectionSelection}
\begin{align}
  \Ket{pI|\psi}&=\lambda_{{\cal C}}\Ket{I|\psi} \label{constraintPriceMinMaxSelection}
  & &\text{Price min/max} \\
   \Ket{V_1|\psi}&=\lambda_{{\cal C}}\Ket{V_0|\psi} \label{constraintPriceMovingAverageEqSelection}
   & &\text{Moving average price (\ref{pVDefine}) is equal to the price (\ref{pvDefine})}
\end{align}
\end{subequations}
The optimization with the constraint
(\ref{constraintPriceMinMaxSelection})
is actually the pure dynamic impact approximation of Ref. \cite{ArxivMalyshkinMuse}: price and execution flow operators are assumed to have the
same eigenvectors.
The (\ref{constraintPriceMovingAverageEqSelection}) states,
same as for the aggregated execution flow (\ref{lambdaEq}) below,
selects the states with the moving average price equals the price,
a typical market practitioner point of attention.
The problem (\ref{ImaxConstarinedSelection}) uses
the same input data moments (\ref{inputmoments})
as the problem (\ref{ImaxConstarined}).

\section{\label{ProjectionsPsiH}  The 
  $\Ket{\psi_{I}^{[IH]}}$ variation approach
  to positive and negative $dI/dt$  states separation.
}
The
separation of the states with positive and negative $dI/dt$ can be developed
based on $\Ket{\psi_{I}^{[IH]}}$ variation.
For example,
in the Eq. (\ref*{Mu-rqD1}) of Ref. \cite{ArxivMalyshkinMuse},
the variation of $I$ have been considered\footnote{
  See the class
  \texttt{\seqsplit{com/polytechnik/utils/RayleighQuotient.java}}
  of \href{http://www.ioffe.ru/LNEPS/malyshkin/AMuseOfCashFlowAndLiquidityDeficit.zip}{provided software},
  implementing the calculation of 0-th, 1-st, and 2-nd variations
  of two quadratic forms ratio.
  }:
\begin{align}
  I_{\psi+\delta\psi}&=\frac{\Braket{\psi+\delta\psi|I|\psi+\delta\psi} }
  {\Braket{\psi+\delta\psi|\psi+\delta\psi}} = D0+D1+D2 +\dots \label{psivar} \\
  D0&=\frac{\Braket{\psi|I|\psi}}{\Braket{\psi|\psi}} \\
  D1&=2\left(\frac{\Braket{\psi|I|\delta\psi}}{\Braket{\psi|\psi}}-
  D0\frac{\Braket{\psi|\delta\psi}}{\Braket{\psi|\psi}}\right) \label{rqD1} \\
  D2&= \frac{\Braket{\delta \psi|I|\delta\psi}}{\Braket{\psi|\psi}}-
  D0\frac{\Braket{\delta \psi|\delta\psi}}{\Braket{\psi|\psi}}
  -2\frac{\Braket{\psi|\delta\psi}}{\Braket{\psi|\psi}}D1 \label{rqD2}
\end{align}
With $\delta \psi=-D(\psi_{I}^{[IH]}(x))$
variation (such a variation can be considered as a boundary condition alternative to (\ref{BOUNDARY:IeqIH}),
  (\ref{BOUNDARY:DIeq0}),
  (\ref{BOUNDARY:IeqI0}),
  or (\ref{BOUNDARY:Ieq0}))
obtain $\Delta_{\psi}P$ from the Eq. (\ref*{Mu-DPDirectRQ})  of Ref. \cite{ArxivMalyshkinMuse}.
Any first variation (\ref{rqD1})
in a $\Ket{\psi_{I}^{[i]}}$ state is zero,
any second variation (\ref{rqD2}) in the state $\Ket{\psi_{I}^{[IH]}}$
is negative.
The first variation of the $\Ket{\psi_{I}^{[IH]}}$ state
 can be written as $P(x)$ polynomial average:
\begin{eqnarray}
  P(x)&=&2\psi_{I}^{[IH]}(x)\left[
    D(\psi_{I}^{[IH]}(x))
-\Braket{\psi_{I}^{[IH]}|D(\psi_{I}^{[IH]})}\psi_{I}^{[IH]}(x)
\right] \label{pvar} \\
  D1&=&\Braket{I\,P(x)}=0 \label{var1eq0}
\end{eqnarray}
In \cite{ArxivMalyshkinLebesgue},
we have have proved, that
any polynomial $P(x)$ of $2n-2$ degree can be
isomorphly mapped to a linear operator of the dimension $n$, the algorithm is
presented in
the Appendix A of Ref. \cite{ArxivMalyshkinLebesgue}:
\begin{eqnarray}
  \rho(x,y)&=&\sum_{i=0}^{n-1}\lambda^{[i]} \psi^{[i]}(x)  \psi^{[i]}(y)  \label{rho}\\
  P(x)&=& \rho(x,x) \label{rhoPrelation}
\end{eqnarray}
Then the $D1$ can be presented as a superposition
of positive and negative terms:
\begin{eqnarray}
  0=D1&=&\sum\limits_{i:\lambda^{[i]}>0}\lambda^{[i]} \Braket{\psi^{[i]}|I|\psi^{[i]}}
  +\sum\limits_{i:\lambda^{[i]}<0}\lambda^{[i]} \Braket{\psi^{[i]}|I|\psi^{[i]}}
  \label{var1eq0average}
\end{eqnarray}
This way the $P(x)$ average can be split in positive and negative
contributions.
Despite being a $\Ket{\psi_{I}^{[IH]}}$ projection, the eigenvalues
of (\ref{rho}) are typically all non--zero,
and corresponding density matrix is a mixed state:
\begin{subequations}
    \label{rhovarIH}
\begin{eqnarray}
  \|\rho^+\|&=&\sum\limits_{i:\lambda^{[i]}>0} \Ket{\psi^{[i]}}\lambda^{[i]}
  \Bra{\psi^{[i]}} \\
  \|\rho^-\|&=&\sum\limits_{i:\lambda^{[i]}<0} \Ket{\psi^{[i]}}\lambda^{[i]}
  \Bra{\psi^{[i]}}
\end{eqnarray}
\end{subequations}
For computer implementation see the class \texttt{\seqsplit{com/polytechnik/trading/DIselDM.java}}
of \href{http://www.ioffe.ru/LNEPS/malyshkin/AMuseOfCashFlowAndLiquidityDeficit.zip}{provided software}.

\bibliography{LD}

\begin{thebibliography}{9}%
\makeatletter
\providecommand \@ifxundefined [1]{%
 \@ifx{#1\undefined}
}%
\providecommand \@ifnum [1]{%
 \ifnum #1\expandafter \@firstoftwo
 \else \expandafter \@secondoftwo
 \fi
}%
\providecommand \@ifx [1]{%
 \ifx #1\expandafter \@firstoftwo
 \else \expandafter \@secondoftwo
 \fi
}%
\providecommand \natexlab [1]{#1}%
\providecommand \enquote  [1]{``#1''}%
\providecommand \bibnamefont  [1]{#1}%
\providecommand \bibfnamefont [1]{#1}%
\providecommand \citenamefont [1]{#1}%
\providecommand \href@noop [0]{\@secondoftwo}%
\providecommand \href [0]{\begingroup \@sanitize@url \@href}%
\providecommand \@href[1]{\@@startlink{#1}\@@href}%
\providecommand \@@href[1]{\endgroup#1\@@endlink}%
\providecommand \@sanitize@url [0]{\catcode `\\12\catcode `\$12\catcode
  `\&12\catcode `\#12\catcode `\^12\catcode `\_12\catcode `\%12\relax}%
\providecommand \@@startlink[1]{}%
\providecommand \@@endlink[0]{}%
\providecommand \url  [0]{\begingroup\@sanitize@url \@url }%
\providecommand \@url [1]{\endgroup\@href {#1}{\urlprefix }}%
\providecommand \urlprefix  [0]{URL }%
\providecommand \Eprint [0]{\href }%
\providecommand \doibase [0]{http://dx.doi.org/}%
\providecommand \selectlanguage [0]{\@gobble}%
\providecommand \bibinfo  [0]{\@secondoftwo}%
\providecommand \bibfield  [0]{\@secondoftwo}%
\providecommand \translation [1]{[#1]}%
\providecommand \BibitemOpen [0]{}%
\providecommand \bibitemStop [0]{}%
\providecommand \bibitemNoStop [0]{.\EOS\space}%
\providecommand \EOS [0]{\spacefactor3000\relax}%
\providecommand \BibitemShut  [1]{\csname bibitem#1\endcsname}%
\let\auto@bib@innerbib\@empty
\bibitem [{\citenamefont {Malyshkin}\ and\ \citenamefont
  {Bakhramov}(2015)}]{2015arXiv151005510G}%
  \BibitemOpen
  \bibfield  {author} {\bibinfo {author} {\bibfnamefont
  {Vladislav~Gennadievich}\ \bibnamefont {Malyshkin}}\ and\ \bibinfo {author}
  {\bibfnamefont {Ray}\ \bibnamefont {Bakhramov}},\ }\bibfield  {title}
  {\enquote {\bibinfo {title} {{Mathematical Foundations of Realtime Equity
  Trading. Liquidity Deficit and Market Dynamics. Automated Trading
  Machines.}}}\ }\href {http://arxiv.org/abs/1510.05510} {\bibfield  {journal}
  {\bibinfo  {journal} {ArXiv e-prints}\ } (\bibinfo {year} {2015})},\ \bibinfo
  {note} {\url{http://arxiv.org/abs/1510.05510}},\ \Eprint
  {http://arxiv.org/abs/1510.05510} {arXiv:1510.05510 [q-fin.CP]} \BibitemShut
  {NoStop}%
\bibitem [{\citenamefont {Malyshkin}(2016)}]{2016arXiv160204423G}%
  \BibitemOpen
  \bibfield  {author} {\bibinfo {author} {\bibfnamefont
  {Vladislav~Gennadievich}\ \bibnamefont {Malyshkin}},\ }\bibfield  {title}
  {\enquote {\bibinfo {title} {{Market Dynamics. On Supply and Demand
  Concepts}},}\ }\href {http://arxiv.org/abs/1602.04423} {\bibfield  {journal}
  {\bibinfo  {journal} {ArXiv e-prints}\ } (\bibinfo {year} {2016})},\ \bibinfo
  {note} {\url{http://arxiv.org/abs/1602.04423}},\ \Eprint
  {http://arxiv.org/abs/1602.04423} {arXiv:1602.04423} \BibitemShut {NoStop}%
\bibitem [{\citenamefont {Malyshkin}(2017)}]{ArxivMalyshkinMuse}%
  \BibitemOpen
  \bibfield  {author} {\bibinfo {author} {\bibfnamefont
  {Vladislav~Gennadievich}\ \bibnamefont {Malyshkin}},\ }\bibfield  {title}
  {\enquote {\bibinfo {title} {{Market Dynamics. On A Muse Of Cash Flow And
  Liquidity Deficit}},}\ }\href {https://arxiv.org/abs/1709.06759} {\bibfield
  {journal} {\bibinfo  {journal} {ArXiv e-prints}\ } (\bibinfo {year}
  {2017})},\ \Eprint {http://arxiv.org/abs/1709.06759} {arXiv:1709.06759
  [q-fin.TR]} \BibitemShut {NoStop}%
\bibitem [{\citenamefont {Malyshkin}\ and\ \citenamefont
  {Bakhramov}(2016)}]{2016arXiv160305313G}%
  \BibitemOpen
  \bibfield  {author} {\bibinfo {author} {\bibfnamefont
  {Vladislav~Gennadievich}\ \bibnamefont {Malyshkin}}\ and\ \bibinfo {author}
  {\bibfnamefont {Ray}\ \bibnamefont {Bakhramov}},\ }\bibfield  {title}
  {\enquote {\bibinfo {title} {{Market Dynamics vs. Statistics: Limit Order
  Book Example}},}\ }\href {https://arxiv.org/abs/1603.05313} {\bibfield
  {journal} {\bibinfo  {journal} {ArXiv e-prints}\ } (\bibinfo {year}
  {2016})},\ \Eprint {http://arxiv.org/abs/1603.05313} {arXiv:1603.05313
  [q-fin.TR]} \BibitemShut {NoStop}%
\bibitem [{\citenamefont {Bobyl}\ \emph {et~al.}(2018)\citenamefont {Bobyl},
  \citenamefont {Davydov}, \citenamefont {Zabrodskii}, \citenamefont {Kostik},
  \citenamefont {Malyshkin}, \citenamefont {Novikova}, \citenamefont
  {Urishov},\ and\ \citenamefont {Yusupova}}]{malyshkin2018spikes}%
  \BibitemOpen
  \bibfield  {author} {\bibinfo {author} {\bibfnamefont {A.V.}\ \bibnamefont
  {Bobyl}}, \bibinfo {author} {\bibfnamefont {V.V.}\ \bibnamefont {Davydov}},
  \bibinfo {author} {\bibfnamefont {A.G.}\ \bibnamefont {Zabrodskii}}, \bibinfo
  {author} {\bibfnamefont {N.R.}\ \bibnamefont {Kostik}}, \bibinfo {author}
  {\bibfnamefont {V.G.}\ \bibnamefont {Malyshkin}}, \bibinfo {author}
  {\bibfnamefont {O.V.}\ \bibnamefont {Novikova}}, \bibinfo {author}
  {\bibfnamefont {D.M.}\ \bibnamefont {Urishov}}, \ and\ \bibinfo {author}
  {\bibfnamefont {E.A.}\ \bibnamefont {Yusupova}},\ }\bibfield  {title}
  {\enquote {\bibinfo {title} {{The Spectral approach to timeserie bursts
  analysis (Спектральный подход к анализу
  всплесков временной
  последовательности)}},}\ }\href
  {https://www.researchgate.net/publication/324171914} {\bibfield  {journal}
  {\bibinfo  {journal} {ISSN 0131-5226.Теоретический и
  научно-практический журнал. ИАЭП.}\ ,\ \bibinfo
  {pages} {77--85}} (\bibinfo {year} {2018})},\ \bibinfo {note}
  {\href{http://dx.doi.org/10.24411/0131-5226-2018-10010}{doi:10.24411/0131-5226-2018-10010}}\BibitemShut
  {NoStop}%
\bibitem [{\citenamefont {Malyshkin}(2018)}]{ArxivMalyshkinLebesgue}%
  \BibitemOpen
  \bibfield  {author} {\bibinfo {author} {\bibfnamefont
  {Vladislav~Gennadievich}\ \bibnamefont {Malyshkin}},\ }\bibfield  {title}
  {\enquote {\bibinfo {title} {{On Lebesgue Integral Quadrature}},}\ }\href
  {https://arxiv.org/abs/1807.06007} {\bibfield  {journal} {\bibinfo  {journal}
  {ArXiv e-prints}\ } (\bibinfo {year} {2018})},\ \Eprint
  {http://arxiv.org/abs/1807.06007} {arXiv:1807.06007 [math.NA]} \BibitemShut
  {NoStop}%
\bibitem [{\citenamefont {Malyshkin}(2014)}]{polynomialcode}%
  \BibitemOpen
  \bibfield  {author} {\bibinfo {author} {\bibfnamefont
  {Vladislav~Gennadievich}\ \bibnamefont {Malyshkin}},\ }\href
  {http://www.ioffe.ru/LNEPS/malyshkin/code.html} {} (\bibinfo {year} {2014}),\
  \bibinfo {note} {the code for polynomials calculation,
  \url{http://www.ioffe.ru/LNEPS/malyshkin/code.html}}\BibitemShut {NoStop}%
\bibitem [{\citenamefont {Bobyl}\ \emph {et~al.}(2016)\citenamefont {Bobyl},
  \citenamefont {Zabrodskii}, \citenamefont {Kompan}, \citenamefont
  {Malyshkin}, \citenamefont {Novikova}, \citenamefont {Terukova},\ and\
  \citenamefont {Agafonov}}]{2016arXiv161107386V}%
  \BibitemOpen
  \bibfield  {author} {\bibinfo {author} {\bibfnamefont
  {Aleksandr~Vasilievich}\ \bibnamefont {Bobyl}}, \bibinfo {author}
  {\bibfnamefont {Andrei~Georgievich}\ \bibnamefont {Zabrodskii}}, \bibinfo
  {author} {\bibfnamefont {Mikhail~Evgenievich}\ \bibnamefont {Kompan}},
  \bibinfo {author} {\bibfnamefont {Vladislav~Gennadievich}\ \bibnamefont
  {Malyshkin}}, \bibinfo {author} {\bibfnamefont {Olga~Valentinovna}\
  \bibnamefont {Novikova}}, \bibinfo {author} {\bibfnamefont
  {Ekaterina~Evgenievna}\ \bibnamefont {Terukova}}, \ and\ \bibinfo {author}
  {\bibfnamefont {Dmitry~Valentinovich}\ \bibnamefont {Agafonov}},\ }\bibfield
  {title} {\enquote {\bibinfo {title} {{Generalized Radon--Nikodym Spectral
  Approach. Application to Relaxation Dynamics Study.}}}\ }\href
  {https://arxiv.org/abs/1611.07386} {\bibfield  {journal} {\bibinfo  {journal}
  {ArXiv e-prints}\ } (\bibinfo {year} {2016})},\ \bibinfo {note}
  {\url{https://arxiv.org/abs/1611.07386}},\ \Eprint
  {http://arxiv.org/abs/1611.07386} {arXiv:1611.07386 [math.NA]} \BibitemShut
  {NoStop}%
\bibitem [{\citenamefont {Golub}(1973)}]{golub1973some}%
  \BibitemOpen
  \bibfield  {author} {\bibinfo {author} {\bibfnamefont {Gene~H}\ \bibnamefont
  {Golub}},\ }\bibfield  {title} {\enquote {\bibinfo {title} {Some modified
  matrix eigenvalue problems},}\ }\href {\doibase 10.1137/1015032} {\bibfield
  {journal} {\bibinfo  {journal} {Siam Review}\ }\textbf {\bibinfo {volume}
  {15}},\ \bibinfo {pages} {318--334} (\bibinfo {year} {1973})}\BibitemShut
  {NoStop}%
\end{thebibliography}%

\makeatletter\@input{am.tex}\makeatother 
\end{document}